\documentclass{nature}

\usepackage{graphicx}
\usepackage[font=small]{caption}
\usepackage{xcolor}
\usepackage{amsmath} 
\usepackage{booktabs}
\usepackage{float} 

\usepackage{multirow}%
\usepackage{amsmath,amssymb,amsfonts}%
\usepackage{amsthm}%
\usepackage{mathrsfs}%
\usepackage[title]{appendix}%
\usepackage{textcomp}%
\usepackage{manyfoot}%
\usepackage{booktabs}%
\usepackage{algorithm}%
\usepackage{algorithmicx}%
\usepackage{algpseudocode}%
\usepackage{listings}%
\usepackage{braket}
\usepackage{url}

\def\bk{{\bf k}}
\def\bq{{\bf q}}
\def\br{{\bf r}}
\def\bR{{\bf R}}
\def\bE{{\bf E}}

\def\hcd{{\hat{c}^\dagger}}
\def\hc{{\hat{c}}}

\title{Ultrafast Strongly Anisotropic Valleytronics in SnSe}

\author
{Yiming Pan$^{1,\dagger}$, Sotirios Fragkos$^{2,\dagger}$, Dominique Descamps$^{2}$, Stéphane Petit$^{2}$, Fabio Caruso$^{1,*}$, Samuel Beaulieu$^{2,\ddag}$}

\begin{document}

\maketitle

\begin{affiliations}
 \item Institut f\"ur Theoretische Physik und Astrophysik, Christian-Albrechts-Universit\"at zu Kiel, 24118 Kiel, Germany
 \item Universit\'e de Bordeaux - CNRS - CEA, CELIA, UMR5107, F33405 Talence, France. \\
  $^{\dagger}$ Equal Contribution \\
  $^{*}$caruso@physik.uni-kiel.de 
  $^{\ddag}$samuel.beaulieu@u-bordeaux.fr
\end{affiliations}

\begin{abstract}
Valleytronics aims to control electrons in a valley-specific manner for quantum information manipulation. Due to their strong in-plane anisotropy, which enables polarization-controlled optical transitions to distinct nondegenerate valleys, group-IV monochalcogenides have been recently proposed as promising candidates for next-generation valleytronic materials. However, ultrafast nonequilibrium dynamics following optical preparation of valley-polarized states remain completely unexplored in these systems. Combining time- and angle-resolved extreme-ultraviolet photoemission spectroscopy with time-dependent Boltzmann equation simulations, we investigate ultrafast valley polarization dynamics following polarization-controlled photoexcitation in SnSe. We show that selective excitation to valleys at global conduction minima yields nearly unity and time-independent valley polarization. In contrast, photoexcitation to the other valley channel leads to ultrafast decay and reversal of valley polarization on sub-picosecond timescales due to intervalley scattering mediated by strong electron-phonon coupling with an optical phonon mode. Our findings reveal strongly anisotropic and radically different nonequilibrium valley physics than in most common two-dimensional valleytronics materials.
\end{abstract}

Valleytronics represents a rapidly emerging frontier that lies at the crossroads of state-of-the-art condensed matter and applied physics, with significant implications for the development of novel quantum information processing devices~\cite{PhysRevLett.108.126804, Rohling_2012, Schaibley16,Xiao12,XiaoDiWang2007}. At its core, valleytronics leverages the valley degree of freedom, i.e., energy extrema of electronic structure in momentum space, offering a new paradigm for quantum information encoding and manipulation. From a fundamental perspective, it provides a versatile platform for investigating valley addressability through polarization-controlled light–matter interactions, where symmetry-guided optical selection rules enable selective excitation of specific valleys~\cite{Yao08, Mak12, Zeng12, Cao12, sciadv.adk3897, Fragkos2025_NC}. Additionally, following the preparation of valley-polarized excited carriers, a plethora of nonequilibrium microscopic processes take place and govern the ultrafast dynamics of valley depolarization~\cite{Zhu25, Yu2014_DiracCones, YuWu2014_ValleyDepolarization, Glazov2015_SpinValley, Yu2015_ValleyExcitons, Selig16, Bertoni16, Xu2021, Molina17}. Understanding and potentially controlling nonequilibrium physical phenomena underlying valley depolarization dynamics is crucial for the design and optimization of valleytronic devices.

Semiconducting transition metal dichalcogenides (TMDCs) are widely regarded as the workhorses of valleytronics. In monolayer TMDCs, resonant photoexcitation using circularly polarized light enables valley-specific transitions at the degenerate K or K' valleys, a phenomenon arising from the broken inversion symmetry of the hexagonal lattice~\cite{Yao08, Mak12, Zeng12, Cao12}. Following valley-selective optical excitation, intervalley scattering occurs on ultrafast timescale through electron–hole exchange interactions~\cite{Yu2014_DiracCones, YuWu2014_ValleyDepolarization, Glazov2015_SpinValley, Yu2015_ValleyExcitons,MaiBarrette14} and via coupling with phonons~\cite{Selig16, Molina17, Xu2021,ws2_valley_lifetime,PanCaruso2023,ChenLi2025}. Several strategies have been employed to mitigate valley depolarization. These strategies include, for example, application of strain~\cite{PhysRevResearch.2.033340, An2023}, tuning the pump photon energy~\cite{PhysRevB.110.125420}, stacking in heterostructures~\cite{doi:10.1126/science.aac7820}, external gating~\cite{Zhang2022}, operating at low temperatures~\cite{Zeng12}, and magnetic doping or application of external magnetic fields to lift valley degeneracy~\cite{PhysRevLett.114.037401, PhysRevB.97.041405}.

Another approach to addressing current challenges in TMDCs-based valleytronics is to identify other material classes with radically different properties and associated working principles, as recently demonstrated in, e.g., silicon and diamond~\cite{Gindl25}. Group-IV tin monochalcogenides (SnX, X=S, Se)~\cite{Barraza21} have recently been proposed as a completely novel valleytronic platform~\cite{Rodin16, Lin2018, Chen18, Tien24, hashmi25, Hanakata2016}. SnX crystallizes in the \textit{Pnma} structure characterized by pronounced anisotropy between the armchair (AC) and zigzag (ZZ) crystal directions. This anisotropy leads to a set of non-degenerate valleys along the ZZ (X valleys -- conduction band minimum) and AC (Y valleys -- valence band maximum) directions, as schematically shown in Fig.~\ref{fig:sketch}(b). Due to their different symmetries and orbital characters, X and Y valleys can be selectively excited using linearly polarized light aligned along ZZ or AC directions, respectively~\cite{Rodin16, Lin2018, Chen18, Tien24, Tolloczko25}. These valley addressability properties have been predicted theoretically~\cite{Rodin16, Tien24} and demonstrated experimentally using optical absorption and emission spectroscopies~\cite{Lin2018, Chen18, Tolloczko25}. These studies report gigantic optical dichroic anisotropies and valley polarization degrees. However, such static optical techniques are sensitive only to direct vertical interband transitions. They reveal valley addressability but fail to capture the ultrafast evolution of carriers in momentum space following photoexcitation. This missing piece is crucial, since valley-based quantum information remains nonvolatile only over the timescale set by intervalley scattering processes, which ultimately erase the optically prepared valley polarization. 

In this work, we combined time- and angle-resolved extreme-ultraviolet photoemission spectroscopy with time-dependent Boltzmann equation (TDBE) simulations to investigate, for the first time, the ultrafast carrier dynamics following polarization-controlled photoexcitation in SnSe. Our experimental measurements reveal that selective excitation to X valleys (ZZ pump), where conduction band minima reside, yields nearly unity and time-independent valley polarization. On the contrary, switching the polarization of the pump field to the AC direction allows for photoexciting electrons within the Y valleys and leads to an ultrafast decay and even a reversal of the valley polarization on sub-picosecond timescales. This highly anisotropic valley polarization dynamics is fully captured by our TDBE simulations, which allows us to identify the main source of depolarization following the optical preparation of X valley polarized states. Indeed, we identified strong electron-phonon scattering with an optical phonon branch as the main driver for Y$\rightarrow$X intervalley scattering. Our results uncover previously unexplored nonequilibrium valley physics in group-IV monochalcogenides, establishing these materials as a new valleytronic platform with distinctive anisotropic behavior, namely, the coexistence of one nearly nonvolatile (time-independent) valley channel and one highly volatile (ultrashort-lived) channel within the same material.

\section{Results and Discussions}

\begin{figure}[H]
\begin{center}
\includegraphics[width=0.65\textwidth]{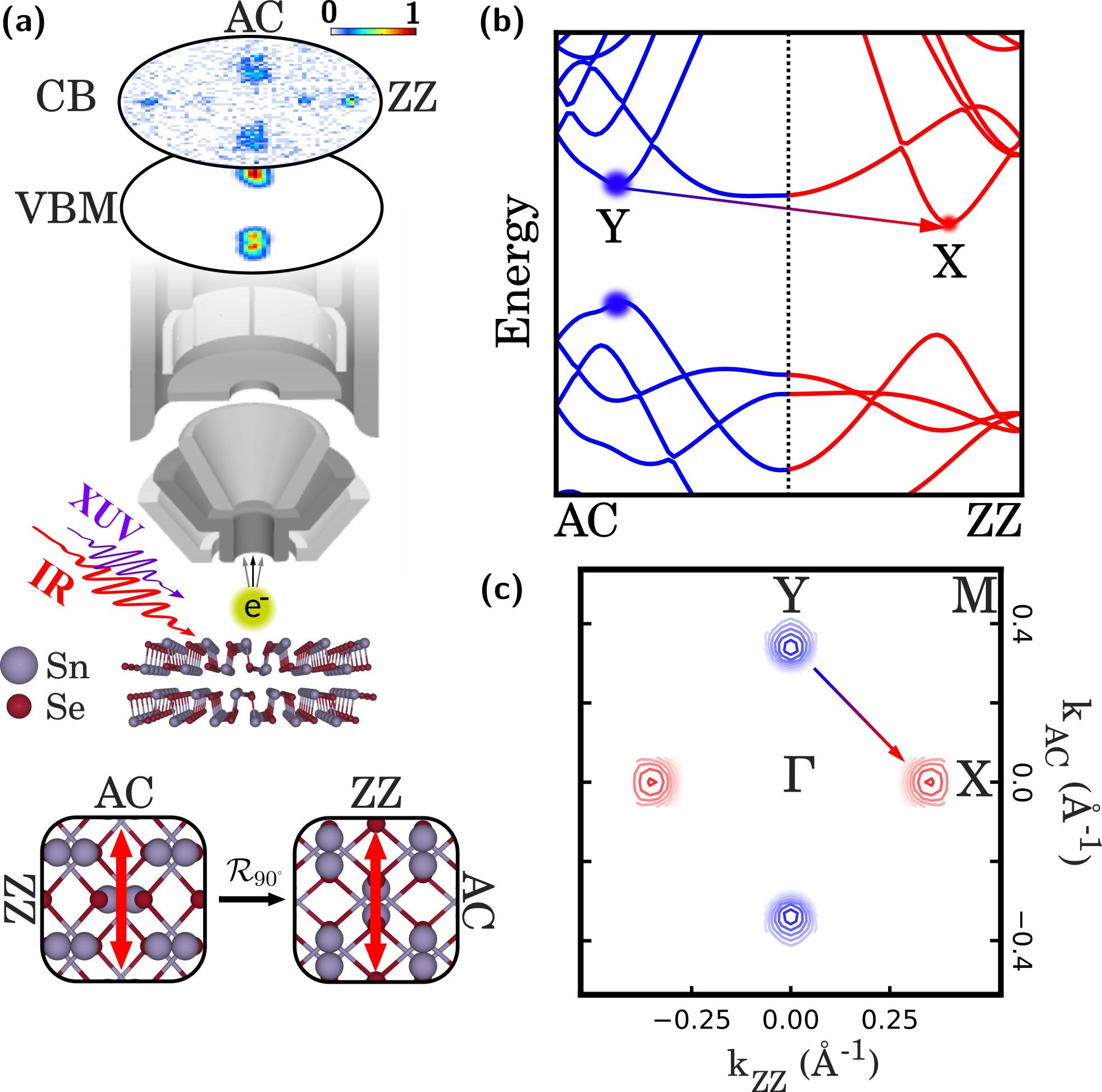}
\caption{\textbf{Scheme of the experiment and ultrafast valleytronics in SnSe}. \textbf{(a)} Schematic of the experimental setup, where s-polarized IR pump pulses (1.2~eV, 135~fs, 0.9~mJ/cm$^2$) and XUV probe pulses (21.6~eV) are sent onto a SnSe crystal along either the AC or ZZ crystallographic axis, with an angle of incidence of 65$^{\circ}$. Time-, energy-, and momentum-resolved photoemission intensities are measured using a time-of-flight momentum microscope. Constant energy contours are shown for energies corresponding to the valence band maximum (VBM: $\mathrm{E-E_{VBM} = 0.00 \pm 0.02~eV}$) and the conduction band (CB: $\mathrm{E-E_{VBM} = 1.15 \pm 0.18~eV}$) for the AC pump configuration. The lower subpanel depicts the real-space in-plane structure of SnSe with pump polarization along ZZ and AC directions. \textbf{(b)} Schematic of SnSe band structure along the AC ($\Gamma$-Y) and ZZ ($\Gamma$-X) high-symmetry directions. \textbf{(c)} Calculated contour plots near the CBM, showing both the X (red) and Y (blue) valleys. The color mapping of each contour progressively shifts toward lighter (whiter) shades with increasing energy. In \textbf{(b)-(c)}, the red-blue arrows indicate the quasi-unidirectional intervalley scattering channel from Y to X valley, upon AC pumping.}
\label{fig:sketch}
\end{center}
\end{figure}

From an experimental point-of-view, our instrument~\cite{Fragkos2025} features a polarization-tunable ultrafast XUV beamline~\cite{Comby22} with a time-of-flight momentum microscope~\cite{Medjanik17, tkach24, tkach24-2} (see Fig.~\ref{fig:sketch}(a)). An s-polarized IR pump pulse (1.2~eV, 135~fs, 0.9~mJ/cm$^2$) and XUV probe pulse (21.6~eV) are collinearly recombined and focused onto a freshly cleaved SnSe (at an angle of incidence (AOI) of 65$^{\circ}$), inside the interaction chamber of the momentum microscope. The sample is held at room temperature, and its alignment and azimuthal orientation are adjusted by a hexapod, allowing for precise orientation between the pump polarization axis and the AC and ZZ crystal axis. More details about the time- and polarization-resolved XUV momentum microscope apparatus can be found elsewhere~\cite{Fragkos2025} and in the Methods section. 

Concerning our theoretical studies, the excitation and relaxation dynamics of carriers following a laser pulse polarized along different directions are investigated by solving the TDBE in real time~\cite{caruso2022, caruso21}. The photoexcitation dynamics are considered via the electronic density matrix $\rho_{mn\bk}$. Carrier relaxation accounts for all relevant electron–phonon scattering pathways in the computation of collision integrals. Further details are presented in the Methods section. For all calculations, we used the same pump pulse duration and intensity as in the experiments, treating the linear polarization direction of the field as the only adjustable parameter. The pump photon energy was chosen to be quasi-resonant with the band gaps at the X and Y valleys obtained from density functional theory (DFT) calculations (0.8 eV), which are known to underestimate band gaps. As discussed below, a slight variation in the pump photon energy is anyway not critical, since valley addressability and subsequent ultrafast carrier dynamics are relatively insensitive to moderate detuning near the band edge, as demonstrated in Figs. S1 and S2 of the Supplementary Materials~\cite{Suppl}.

\begin{figure}[H]
\begin{center}
\includegraphics[width=\textwidth]{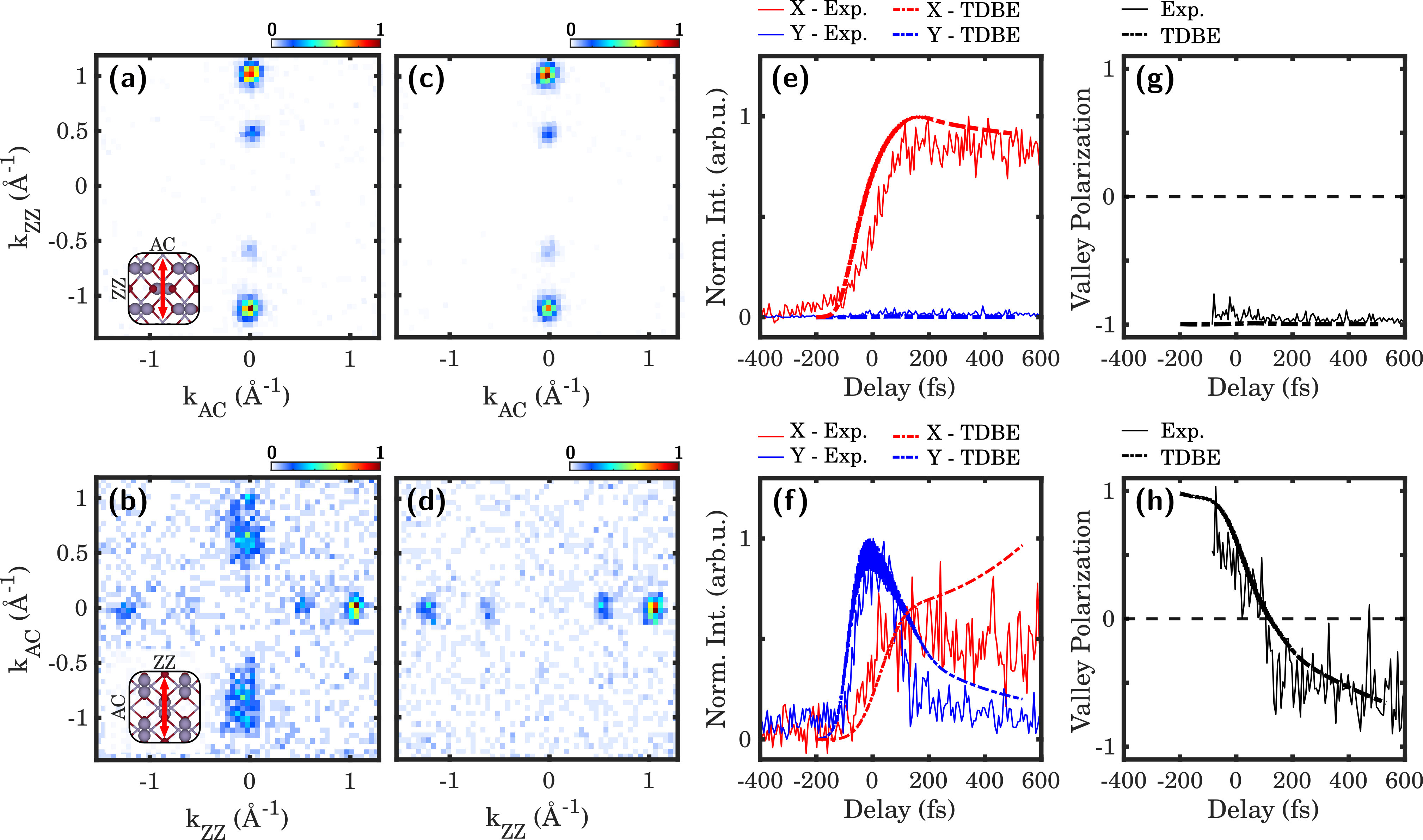}
\caption{\textbf{Ultrafast strongly anisotropic valley-resolved electron dynamics in SnSe}. \textbf{(a)-(d)} Momentum distribution of excited electrons within conduction bands, in \textbf{(a),(c)} and \textbf{(b),(d)} following photoexcitation using IR polarization along ZZ and AC direction, respectively. \textbf{(a),(b)} and \textbf{(c),(d)} are measured at early (-125~fs to +125~fs) and late (+350~fs to +600~fs) time delays after polarization-controlled photoexcitation. \textbf{(e),(f)} Experimental and calculated valley-resolved electron dynamics following photoexcitation using IR polarization along ZZ and AC direction, respectively, and \textbf{(g),(h)} associated valley polarization dynamics extracted by taking the normalized difference between intensity within CB X and Y valleys shown in \textbf{(e)} and \textbf{(f)}, respectively.}
\label{fig2}
\end{center}
\end{figure}

We start by examining the initial momentum distribution of excited carriers within the conduction band (CB) following photoexcitation by an IR pump polarized along the ZZ and AC directions (Fig.~\ref{fig2}(a) and (b), respectively). Due to the non-normal geometry of our photoemission experiments (angle of incidence, AOI = 65$^{\circ}$), rotating the azimuthal angle of the sample and rotating the polarization axis of the IR pump pulses are not equivalent operations. In particular, keeping the crystal orientation fixed and switching from s- to p-polarized light to photoexcite along the ZZ and AC directions, although experimentally simpler, would introduce spurious effects arising from strong laser-assisted photoemission. To avoid these detrimental effects, we fix the pump polarization (s-polarization, i.e., in-plane) and azimuthally rotate the crystal so that the ZZ or AC direction aligns with the pump polarization axis. 

When the polarization axis is aligned with the ZZ crystal axis (ZZ pumping), it has been both predicted~\cite{Rodin16} and experimentally confirmed via polarization-resolved optical spectroscopies~\cite{Lin2018, Chen18, Tien24, Tolloczko25} that optical interband transitions occur exclusively around the X valleys, which constitute the global conduction band minimum (CBM). Our measurements are consistent with these optical selection rules, showing that the momentum distributions observed at both early and late delays after ZZ pumping (Fig.~\ref{fig2}(a) and (c)) are nearly identical and exhibit a remarkably simple pattern. In this light–matter interaction geometry, only the X valleys of the CB are populated, and the carrier population remains strongly localized within these valleys even at longer pump–probe delays. This behavior is further underscored by the ultrafast valley-resolved carrier dynamics (Fig.~\ref{fig2}(e)) and by the corresponding valley polarization dynamics (Fig.~\ref{fig2}(g)). Specifically, we observe a rapid rise of the photoemission intensity from the CB X valleys (red curve), while the Y valleys show no significant signal (blue curve), yielding a nearly time-independent and almost fully valley-polarized population (Fig.~\ref{fig2}(g)). This behavior is fully captured by our TDBE simulations, where the population of the electrons in different valleys is obtained by taking the diagonal elements of the electronic density matrix, \textit{i.e.} $f_{n\bk}(t) = \rho_{nn\bk}(t)$ averaged for momentum $\bk$ around X- and Y-valleys.

Going from ZZ to AC pumping by azimuthally rotating the crystal by 90$^{\circ}$ produces strikingly different results (Fig.~\ref{fig2}(b,d,f,h)). Indeed, upon AC pumping, optical interband transitions occur around the Y valleys~\cite{Rodin16, Lin2018, Chen18, Tien24, Tolloczko25}, which correspond to a \textit{local} conduction band minimum. At early time delays following AC photoexcitation (Fig.~\ref{fig2}(b)), significant photoemission intensity emerges from the CB Y valleys, a feature completely absent under ZZ pumping. Additionally, a nonvanishing photoemission intensity is also observed at the CB X valleys, which can potentially originate from ultrafast intervalley scattering during the pump pulse duration. At later pump-probe delays (Fig.~\ref{fig2}(d)), the vast majority of the CB population resides at the global CBM, i.e., within the X valleys. This momentum-space population redistribution strongly suggests intervalley scattering from the optically prepared Y valleys to the global CBM at the X valleys. This mechanism is further illustrated by the ultrafast valley-resolved carrier dynamics (Fig.~\ref{fig2}(f)) and the corresponding valley polarization dynamics (Fig.~\ref{fig2}(h)). Specifically, under AC pumping, we observe a rapid rise and fast decay of the photoemission intensity from the CB Y valleys (blue curve), accompanied by a delayed rise and slower decay of the intensity in the X valleys (blue curve). These ultrafast valley-dependent carrier dynamics result in an initially strong positive valley polarization immediately following the interband optical transition (pump pulse), which subsequently decays and even reverses over time (Fig.~\ref{fig2}(h)). In Fig.~\ref{fig2}(f), our TDBE simulations accurately reproduce the early-time dynamics at both the X and Y valleys. However, at longer time delays (i.e., between 200~fs and 600~fs), a continuous increase in the X-valley population is observed. This feature, absent in the experimental data, likely arises from the neglect of other relaxation mechanisms, such as electron–hole recombination, in our theoretical model. Nevertheless, as shown in Fig.~\ref{fig2}(f), the TDBE-predicted valley depolarization dynamics are in excellent agreement with the experimental results. Our TDBE simulations also enable tracking the real-time dynamics of photoholes. The results reveal a polarization-dependent valley behavior opposite to that of electrons: light-induced holes undergo an ultrafast valley polarization reversal under ZZ pumping, while their polarization remains constant under AC pumping. These findings are discussed extensively in the Supplementary Materials~\cite{Suppl} (see Fig. S3).

Many important conclusions can already be drawn from the time-, polarization-, and valley-resolved carrier dynamics presented in Fig.~\ref{fig2}. Our measurements reveal that even after optically establishing strong and opposite valley polarization under AC and ZZ photoexcitation, subsequent microscopic coupling mechanisms induce unidirectional intervalley scattering toward the global CBM, effectively reversing the optically prepared valley polarization on a timescale of hundreds of femtoseconds following AC pumping. This highly anisotropic behavior, i.e., the coexistence of both highly volatile (fast decay and reversal) and nonvolatile (fully valley polarized and time-independent) channels, arising from valley nondegeneracy, is fundamentally different from standard TMDC-based valleytronics. Indeed, under similar conditions (room temperature), valley degeneracy and strong intervalley coupling in, for example, a WS$_2$ monolayer, lead to symmetric erasure of the initial valley polarization on a sub-picosecond timescale for both valley channels (K and K')~\cite{Kunin23,Zhu25}. This highlights that group-IV monochalcogenides and TMDCs exhibit fundamentally different nonequilibrium valley physics, suggesting potentially distinct functionalities in valleytronic devices. 

\begin{figure}[H]
\begin{center}
\includegraphics[width=0.8\textwidth]{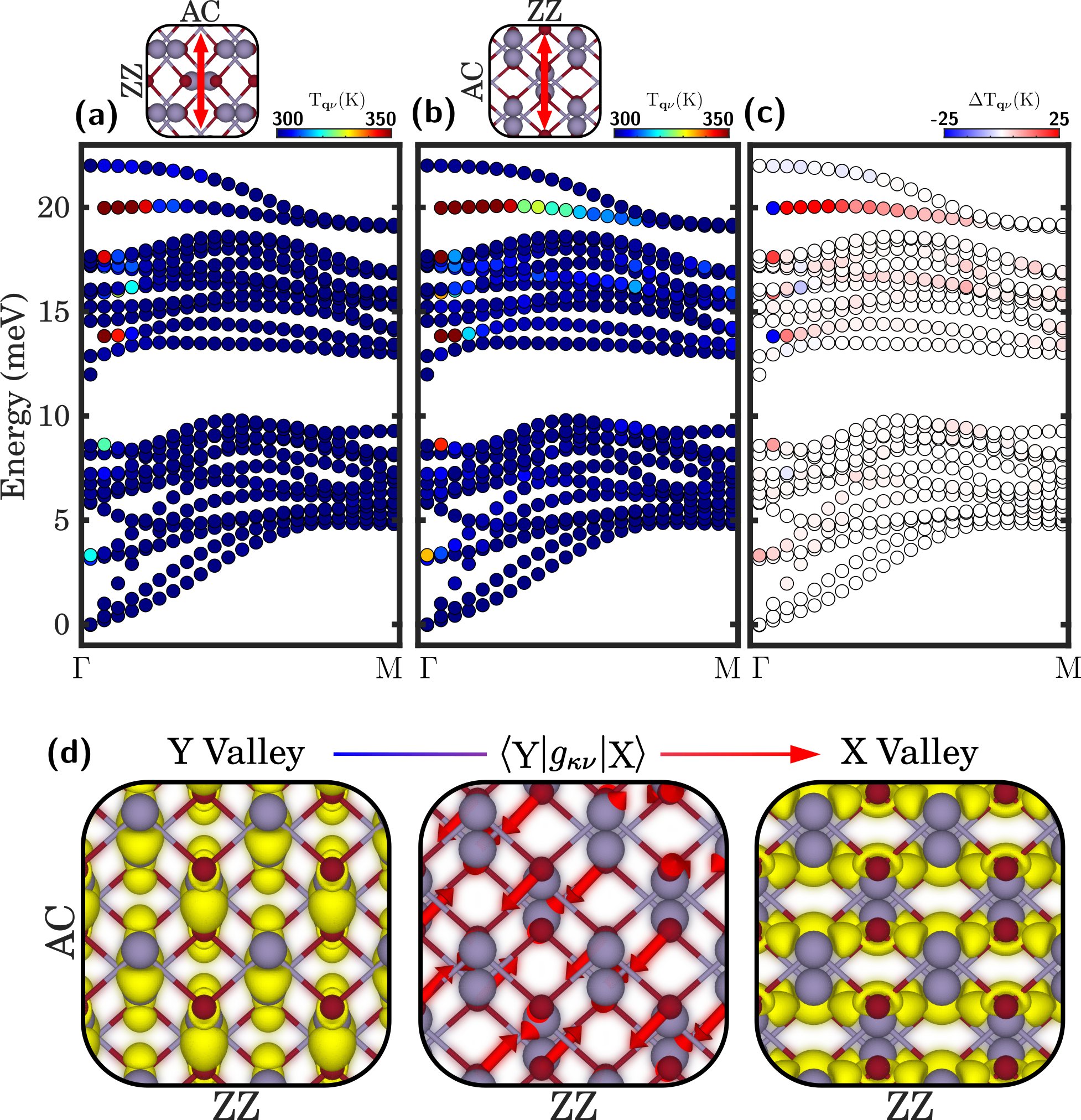}
\caption{\textbf{Microscopic mechanism underlying ultrafast valley depolarization}. \textbf{(a)–(c)} Phonon dispersion and momentum- and mode-resolved phonon temperatures along the $\Gamma$–M high-symmetry direction, at the end of the TDBE simulations (600~fs after the pump), following photoexcitation along the ZZ and AC directions, respectively. \textbf{(c)} Mode-resolved differential phonon temperature, obtained by subtracting the phonon temperature under AC \textbf{(b)} and ZZ \textbf{(a)} pumping. \textbf{(d)} Analysis of the phonon mode responsible for Y$\rightarrow$X intervalley scattering. The left and right panels show the electronic wave functions at the bottom of the conduction band at the Y and X valleys, respectively. The middle panel illustrates the atomic displacements (red arrows) associated with the strongly coupled optical phonons around 20~meV, which drive the Y$\rightarrow$X intervalley scattering.}
\label{fig:ph_dynamics}
\end{center}
\end{figure}

To go one step further in the understanding of the mechanism underlying unidirectional Y$\rightarrow$X intervalley scattering, we investigate quantum-state-resolved and mode-specific electron–phonon interaction following polarization-controlled optical preparation of valley-polarized states and subsequent dynamics driven by electron-phonon coupling. To do so, we extracted the energy- and momentum-resolved nonequilibrium phonon distribution 600 fs after the pump pulse, along the high-symmetry path $\Gamma$-M (Fig.~\ref{fig:ph_dynamics}). Indeed, due to momentum conservation, the intervalley scattering of carriers leads to the emission of phonons along the $\Gamma$-M direction. By comparing the momentum- and mode-resolved phonon temperature, defined as $T_{\bq \nu} =\hbar \omega_{\bq\nu}
[{k_{\rm B}{\rm ln}(1+n_{\bq \nu}^{-1})}]^{-1}$, where $k_{\rm B}$, $\omega_{\bq\nu}$ and $n_{\bq \nu}$ are respectively Boltzmann constant, phonon frequency and phonon population, we can unambiguously identify which phonon modes dominantly contribute to electron-phonon coupling. In Fig.~\ref{fig:ph_dynamics}(a), upon ZZ pumping, only optical modes around $\Gamma$ are populated, since only intravalley scattering is possible. For AC pumping, phonons are emitted close to $\Gamma$ and at finite momentum along $\Gamma$-M, as shown in Fig.~\ref{fig:ph_dynamics}(b), due to the presence of both intravalley and intervalley scattering channels. We further highlight the difference by computing the difference $\Delta T_{\bq\nu}=T^{\rm AC}_{\bq\nu}-T^{\rm ZZ}_{\bq\nu}$ for the two photoexcitation scenarios in Fig.~\ref{fig:ph_dynamics}(c), where we can see that the second highest optical branch ($\sim$20~meV) is responsible for the intervalley scattering. This phonon mode has been previously shown to dominate electron-phonon coupling and to play a decisive role in thermoelectric properties of SnSe~\cite{CarusoTroppenz2018}. The mode-selective phonon emission is decided by the e-ph coupling matrix elements $g_{mn\nu}(\bk,\bq) = \frac{1}{\sqrt{2\omega_{\bq\nu}}}\bra{m\bk+\bq}\partial_{\bq\nu}V^{\rm scf}\ket{n\bk}$, defined as the overlap between the two electronic wave functions by phonon-induced potential. Thereby, we performed symmetry analysis on both the wave functions of the X-valley and Y-valley. They are both even under the gliding symmetry on the x-y plane ($\bar{M}_{xy}$), hence the phonon connecting these two states should also be even under $\bar{M}_{xy}$, which is satisfied by the second-highest optical branch~\cite{Suppl}. The electronic states at X- and Y-valleys are essentially made of $p_x$ and $p_y$ orbitals of the Sn atoms. The large in-plane displacement of the optical phonon, as shown in Fig.~\ref{fig:ph_dynamics}(d), plays the role of connecting these two orthogonal orbitals, making the corresponding e-ph coupling matrix elements nonzero. We verify this by projecting the e-ph coupling matrix element that connects the two valley states at X and Y into atomic displacements (see Fig. S4 of SM~\cite{Suppl}), which suggest that in-plane atomic motions contribute largely to the intervalley scatterings. Having established SnSe as a novel singular valleytronic platform, our state-resolved electron–phonon coupling analysis based on TDBE enables a direct identification of the microscopic mechanisms that drive its valley depolarization.

\section*{Conclusions}
In conclusion, by combining time- and angle-resolved extreme-ultraviolet photoemission spectroscopy with state-of-the-art TDBE simulations, we report the first observation of ultrafast valleytronics in group-IV monochalcogenides and uncover an operating principle that is fundamentally distinct from all existing valleytronic platforms. The pronounced in-plane anisotropy and resulting nondegenerate valleys in SnSe give rise to ultrafast and strongly anisotropic valley dynamics, featuring one nearly nonvolatile (time-independent) and one highly volatile (ultrashort-lived) valley channel within the same material. Our TDBE simulations reveal that the ultrafast anisotropic valley depolarization originates from a quantum-state-resolved and mode-specific electron–phonon interaction, enabling us to gain a holistic understanding of nonequilibrium valley physics in this new valleytronic platform. The mechanism of phonon-driven anisotropic valleytronics with coexisting volatile and nonvolatile channels is expected to be general and applicable to any system with addressable nondegenerate valleys. Our work thus opens promising avenues for future theoretical and experimental studies across a broad class of materials with broken valley degeneracy, including magnetically doped TMDCs~\cite{Macneill15, Qi15}, van der Waals heterostructures~\cite{Nagler17}, twisted altermagnets~\cite{Guo24}, and ferrovalley materials~\cite{Tong2016}. The discovery of this strongly anisotropic valleytronic platform opens new perspectives for valley-based quantum and optoelectronic functionalities.

\section*{Methods}
\subsection{Experimental setup}\label{sec:exp}

The experimental setup revolves around a polarization-tunable ultrafast extreme ultraviolet (XUV) beamline~\cite{Comby22} coupled to a momentum microscope end-station~\cite{tkach24, tkach24-2, Fragkos2025}. We use a high-repetition-rate Yb fiber laser (166~kHz, 1030~nm, 135~fs, 50~W, Amplitude Laser Group). For the HHG-based XUV probe arm, a significant portion of the laser beam is frequency-doubled in a BBO crystal to generate a few watts of 515~nm pulses. After beam size adjustment and spatial profile tailoring to create an annular beam, these pulses are focused into an argon gas jet to produce XUV radiation through high-order harmonic generation. To reject the 515~nm driving beam from the XUV beamlet, we use sequential spatial filtering using pinholes. The 9th harmonic of the 515~nm driver (21.6 eV) is spectrally isolated via reflections onto Sc/SiC multilayer XUV mirrors (NTTAT) and propagation through a 200~nm thick Sn filter (Luxel). For the IR pump arm, a small fraction of the fundamental laser pulse (1030~nm, 135~fs) is utilized. The IR pump and XUV probe pulses are collinearly recombined using a drilled mirror and are subsequently focused onto the sample, achieving typical spot sizes of 70 $\mu$m $\times$ 140 $\mu$m for the IR pump and 45 $\mu$m $\times$ 35 $\mu$m for the XUV probe, respectively. Concerning sample preparation, bulk $\mathrm{SnSe}$ samples (HQ Graphene) are glued on the sample holder (Flag style - Ferrovac), and a cylindrical ceramic post (Umicore) is glued on the top surface of the sample using UHV-compatible conductive epoxy (Epo-tek). The samples are then cleaved by hitting the ceramic post in an ultrahigh vacuum environment at a base pressure of 1$\times$10$^{-10}$ mbar. The freshly cleaved sample is then introduced into a motorized hexapod for precise sample alignment within the main chamber (base pressure: 2$\times$10$^{-10}$ mbar, room temperature). Time-resolved photoemission measurements are performed using a custom time-of-flight momentum microscope equipped with an advanced front lens offering multiple operational modes (GST mbH)~\cite{tkach24, tkach24-2}. For data post-processing, we employ an open-source data workflow~\cite{Xian20, Xian19_2} to efficiently transform raw single-event datasets into calibrated, binned hypervolumes of the desired dimensions. More details about the experimental setup can be found in Ref.~\citenum{Fragkos2025}.

\subsection{Time-dependent Boltzmann equation}\label{sec:th}

The real-time dynamics of electrons and phonons are simulated using the time-dependent Boltzmann equation (TDBE), which are a set of equations for the temporal evolution of electron population $f_{n\bk}$ and phonon populations $n_{\bq\nu}$. In Hartree atomic units, these equations can be expressed simply as:
\begin{align}
    \partial_tf_{n\bk} &= \left.\partial f_{n\bk}\right|_{pump} +\Gamma^{\rm ep}_{n\bk} \label{f-ep-relax} \quad, \\
    \partial_tn_{\bq\nu}&=\Gamma_{\bq\nu}^{\rm pe} \label{n-ep-relax}\quad,
\end{align}
where the electronic dynamics are driven by photoexcitation, denoted as $\left.\partial f_{n\bk}\right|_{pump}$ and relaxation process resulting from electron-phonon scatterings, which are described by the electron-phonon collision integrals $\Gamma^{\rm ep}_{n\bk}$. The out-of-equilibrium phonon dynamics is driven by the collision integrals $\Gamma^{\rm pe}_{\bq\nu}$. The expression of collision integrals can be derived from many-body perturbation theory~\cite{KuhnRossi1992}, and are provided in Ref.~\citenum{Caruso22}. The photo-excitation dynamics, denoted as $f_{n\bk}|_{pump}$ is a fully coherent process, and are accounted for via the semiconductor Bloch equation, where the dynamics of the density matrix $\rho_{mn\bk}=\langle\hcd_{m\bk}\hc_{n\bk}\rangle$ are numerically solved:
% \begin{widetext}
\begin{align}
    \left[\partial_t+i\left(\omega_{m\bk}-\omega_{n\bk}-i\frac{1}{\tau_{\rm dph}}(1-\delta_{mn})\right)\right]\rho_{mn\bk}|_{pump} = i\sum_l \bE(t)\cdot\left[\br_{ml\bk}\rho_{ln\bk}-\rho_{ml\bk}\br_{ln\bk}\right] \label{f-pump-coh} \quad,
\end{align}
% \end{widetext}
where $\omega_{n\bk}$ represents the eigenvalue of the electronic states, $\br_{mn\bk}=i\langle u_{m\bk}|\nabla_{\bk}u_{n\bk}\rangle$ is the interband Berry connection, and $\bE(t)$ denotes the linearly-polarized pump field along the direction ${\bf e}$, simulated by the following function:
\begin{align}
    \bE(t) =E_0{\rm sin}(\omega_0 t){\rm exp}\left[-\frac{(t-t_0)^2}{2\sigma^2}\right]{\bf e}\quad,
\end{align}
where $\omega_0$ is the photon energy and $E_0$ is the maximum of electric field. We employed the photon energy of 0.8 eV for the pump field, which matches the direct band gap predicted by DFT calculations. The parameters $E_0$, $t_0$, and $\sigma$ are decided by the pump field employed in the experiment. Eq.~\eqref{f-pump-coh} is coupled to Eq.~\eqref{f-ep-relax} via the relation $f_{n\bk} = \rho_{nn\bk}$. $1/\tau_{\rm dph}$ represents the dephasing rate of the interband coherence, which is related to the imaginary part of the self-energy of electron-hole pair~\cite{KuhnRossi1992}. We employ a phenomenological dephasing time of 20 fs. Finally, the coupled equations, Eqs.~\eqref{f-ep-relax}-\eqref{f-pump-coh} are solved simultaneously in real-time using 4th-order Runge-Kutta method, and are implemented in {\tt EPW} code~\cite{PanCaruso2023,Noffsinger2010,giustino07,Lee2023}.

\subsection{Density Functional Theory Calculations}
The parametrization of TDBE relies on the Density functional theory (DFT) and Density functional perturbation theory (DFPT) calculations for the electron and phonon band structures~\cite{Giannozzi2017, Baroni2001}, as implemented in the \texttt{Quantum Espresso} code~\cite{Giannozzi2017}. We employ norm-conserving Hartwigsen-Goedecker-Hutter pseudopotential~\cite{hgh_pseudopotential} and PBE-GGA approximation to exchange-correlational functions~\cite{PBE_GGA} in our DFT and DFPT calculations. A kinetic energy cutoff of 120 Ry is used to expand the wave functions into plane waves. All calculations employed the DFT-relaxation crystal structures, which predict lattice parameters $a_1=4.226$ \AA, $a_2=4.547$ \AA, and $a_3=11.830$ \AA ~in the three directions of the rectangular cuboid unit cell. The Brillouin zone is sampled with $15 \times 15 \times 5$ homogeneous $\bk$ points. The phonons are computed with DFPT on a $3 \times 3 \times 2$ $\bq$-grid. The e-ph coupling matrix elements are calculated within the \texttt{EPW} code~\cite{Noffsinger2010,giustino07,Lee2023}, where $s$ and $p$ orbitals of Sn and Se are used as initial projectors for constructing maximally localized wannier functions (MLWFs)~\cite{Marzari2012}. The electron energies, phonon frequencies, and e-ph coupling matrix elements are interpolated on a $45\times45\times 10$ homogeneous grid for both $\bk$ and $\bq$ points. For the time derivative in the real-time equations, we use a time-step of 1 femtosecond for the relaxation dynamics Eqs.~\eqref{f-ep-relax}-\eqref{n-ep-relax}, and 5 attoseconds (0.005 fs) for the photoexcitation dynamics Eq.~\eqref{f-pump-coh}. A Gaussian smearing of 3 meV is used for the calculation of the e-ph collision integrals.

\section*{Data Availability}
The data that support the findings of this article will be openly available on Zenodo 
 \url{}. 

\section*{Acknowledgements}
We thank Yann Mairesse for insightful discussions. We thank Nikita Fedorov, Romain Delos, Pierre Hericourt, Rodrigue Bouillaud, Laurent Merzeau, and Frank Blais for technical assistance.  We thank Baptiste Fabre for implementing and maintaining the data binning code. We acknowledge the financial support of the IdEx University of Bordeaux/Grand Research Program "GPR LIGHT". We acknowledge support from ERC Starting Grant ERC-2022-STG No.101076639, Quantum Matter Bordeaux, AAP CNRS Tremplin and AAP SMR from Université de Bordeaux. This work is part of the ULTRAFAST and TORNADO projects of PEPR LUMA and was supported by the French National Research Agency, as a part of the France 2030 program, under grants ANR-23-EXLU-0002 and ANR-23-EXLU-0004. Y.P and F.C acknowledge the funding by Deutsche Forschungsgemeinschaft (DFG) -- project number 443988403. The simulations are performed thanks to the computing time provided by the high-performance computer Lichtenberg at the NHR Centers NHR4CES at TU
Darmstadt (Project p0021280). S.F. acknowledges funding from the European Union’s Horizon Europe research and innovation programme under the Marie Skłodowska-Curie 2024 Postdoctoral Fellowship No 101198277 (TopQMat). Funded by the European Union. Views and opinions expressed are however those of the author(s) only and do not necessarily reflect those of the European Union. Neither the European Union nor the granting authority can be held responsible for them. 

\section*{Author contributions}
S.B. conceived the research project. S.F. and S.B. performed the experiments. S.B. analyzed the experimental data. D.D. and S.P. participated in maintaining the laser system. Y.P. and F.C. developed the theoretical framework. Y.P. performed simulations and numerical implementations. S.B. and Y.P. wrote the first draft of the manuscript. All authors participated in commenting and revising the manuscript. 

\clearpage

\setcounter{figure}{0} 
\renewcommand{\figurename}{Fig.}
\renewcommand{\thefigure}{S\arabic{figure}}

\setcounter{section}{0} 
\renewcommand{\thesection}{S\arabic{section}}

\setcounter{subsection}{0} 
\renewcommand{\thesubsection}{S\arabic{section}.\arabic{subsection}}

\setcounter{equation}{0} 
\renewcommand{\theequation}{S\arabic{equation}}

\setcounter{table}{0}  
\renewcommand{\thetable}{S\arabic{table}}

\begin{center}
    {\Large \textbf{Supplementary information for: \\ Ultrafast Strongly Anisotropic Valleytronics in SnSe}}
\end{center}

\section{Valley- and polarization-resolved optical properties of \NoCaseChange{SnSe}}\label{sec:opt}

\begin{figure}[H]
\begin{center}
\includegraphics[width=0.75\textwidth]{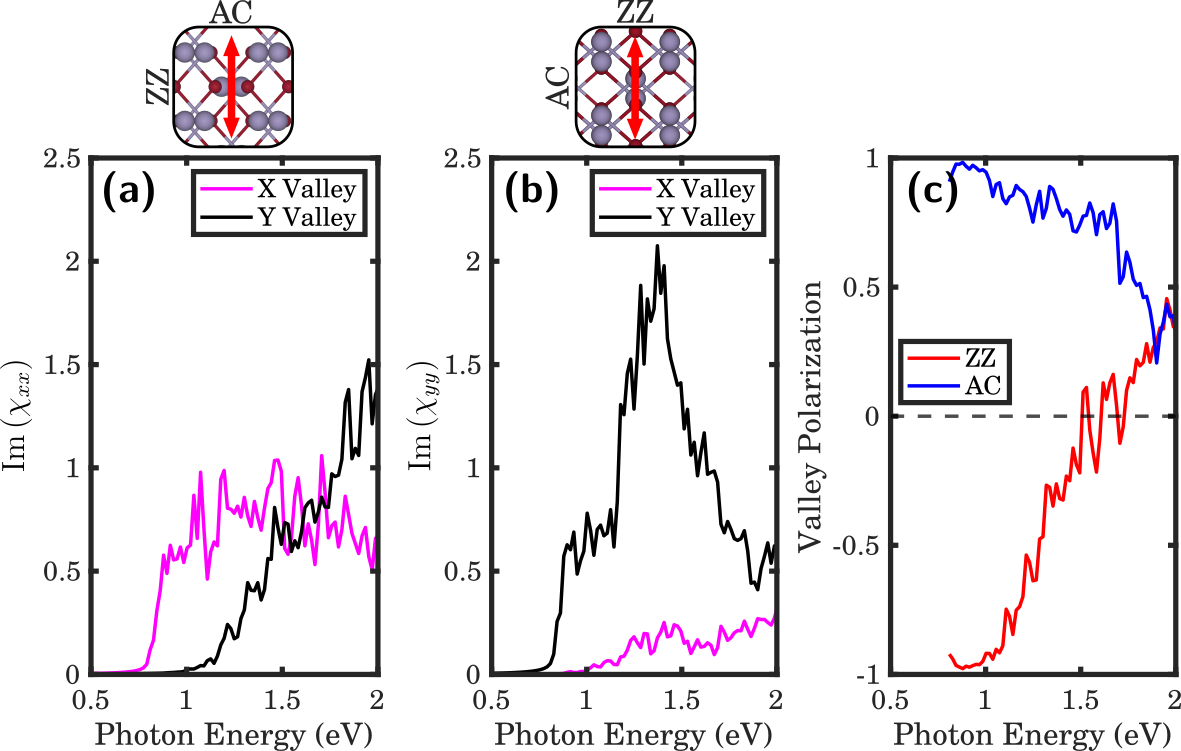}
\caption{\textbf{Valley- and polarization-resolved optical properties of SnSe}. \textbf{(a)-(b)} Photon-energy- and valley-resolved (X-valley in pink and Y-valley in black) imaginary part of the electric susceptibility tensor for light polarization axis along ZZ ($\mathrm{Im}(\chi_{xx})$ in \textbf{(a)} and AC ($\mathrm{Im}(\chi_{yy})$ in \textbf{(b)}) directions. \textbf{(c)} Photon-energy-dependent valley polarization for light polarization axis along ZZ (in red) and AC (in blue) directions.}
\label{fig:absorption}
\end{center}
\end{figure}

To investigate the role of pump photon energy on the initial valley polarization following interband transitions, we have computed the energy-, polarization-, and valley-resolved imaginary part of the electric susceptibility tensor, i.e., $\mathrm{Im}\chi_{aa}$, where $a = x,y,z$.  The imaginary part of the electric susceptibility tensor is proportional to the material’s absorption coefficient, and can be written as (in Hartree atomic units)
\begin{align}
   {\rm Im}\chi_{aa}(\omega) = \frac{4\pi^2}{N_{\bk}V_{uc}}\sum_{cv\bk}2\left|\bra{c\bk}\hat{r}_a\ket{v\bk}\right|^2\delta(\omega_{c\bk}-\omega_{v\bk}-\omega ) \label{eq:absorption}\quad,
\end{align}
where $N_{\bk}$ is the total number of $\bk$ points, $V_{uc}$ is the volume of the unit cell, and the additional factor of 2 accounts for the spin degeneracy. $\hat{r}_a$ represents the position operator in the $a$ ($a=x,y,z$) direction, and the matrix elements is the interband Berry connection, \textit{i.e.} $\bra{c\bk}\hat{r}_a\ket{v\bk}=i\bra{u_{c\bk}}\partial_{k_a}u_{v\bk}\rangle$. From Eq.~\eqref{eq:absorption}, it is known that the absorption for linearly polarized light along $a$ direction is proportional to the joint density of state at photon energy $\omega$, modulated by the interband dipole matrix elements. Since the direct band gaps at X-valley and Y-valley predicted by our DFT calculation are quasi-degenerate (around 0.8 eV), the absorption rate ${\rm Im}\chi_{aa}(\omega)$ for $x$ and $y$ are both nonzero. In view of this, we define valley-polarized absorption rate by limiting the sum over $\bk$ in Eq.~\eqref{eq:absorption} to the regions around X-valley and Y-valley to investigate the contribution of interband excitation at different valleys to the absorption rate. This method was previously employed in WS$_2$ to study the contribution of valley exciton to absorption under circularly-polarized irradiation~\cite{Caruso22}. The energy-resolved strength of $\mathrm{Im}(\chi_{aa})$ at the X- and Y-valleys for pump polarization aligned along the ZZ ($\mathrm{Im}(\chi_{xx})$) and AC ($\mathrm{Im}(\chi_{yy})$) crystal axes (see Figs.~\ref{fig:absorption}(a) and (b)) allows us to theoretically investigate the influence of pump photon energy on the initial valley polarization, i.e., immediately after photoexcitation, before any scattering processes occur.

When the pump photon energy is close to resonance with the DFT band gap ($\sim 0.8~\mathrm{eV}$), the optical selection rules are well defined: pumping along the ZZ (AC) direction leads to exclusive absorption at the X- (Y-) valleys, resulting in near-unity valley polarization. As the photon energy increases, direct vertical interband transitions start to occur away from the band edge, relaxing these selection rules and reducing the valley polarization. Due to the strong anisotropy of the electronic structure of SnSe, the rate at which valley polarization decays with photon energy detuning strongly depends on the pump polarization direction, with polarization decaying faster for the ZZ than the AC configuration. Nevertheless, the main conclusion from this analysis is that detuning up to $\sim 300~\mathrm{meV}$ above the band edge (i.e., from 0.8~eV to 1.1~eV) still leads to near-unity initial valley polarization, as shown in Fig.~\ref{fig:absorption}(c). This confirms that, in our experiment, even though the pump photon energy is slightly above the band edge, the optical selection rules enable the preparation of strongly valley-polarized electrons. 

\section{Role of detuning in real-time electron dynamics in \NoCaseChange{SnSe}}\label{sec:detuning}

After investigating the role of pump photon energy on the initial valley polarization following interband transitions, we turn our attention to the role of detuning on real-time electron dynamics. Indeed, in the previous section, we have shown that moderate detuning of photon energy with respect to the band edge does not alter the preparation of strongly valley-polarized electrons. However, the population of higher-lying energy states within the conduction band upon pumping above the band edge can have some consequences on the subsequent valley-resolved real-time electron dynamics. 

\begin{figure}[H]
\begin{center}
\includegraphics[width=0.75\textwidth]{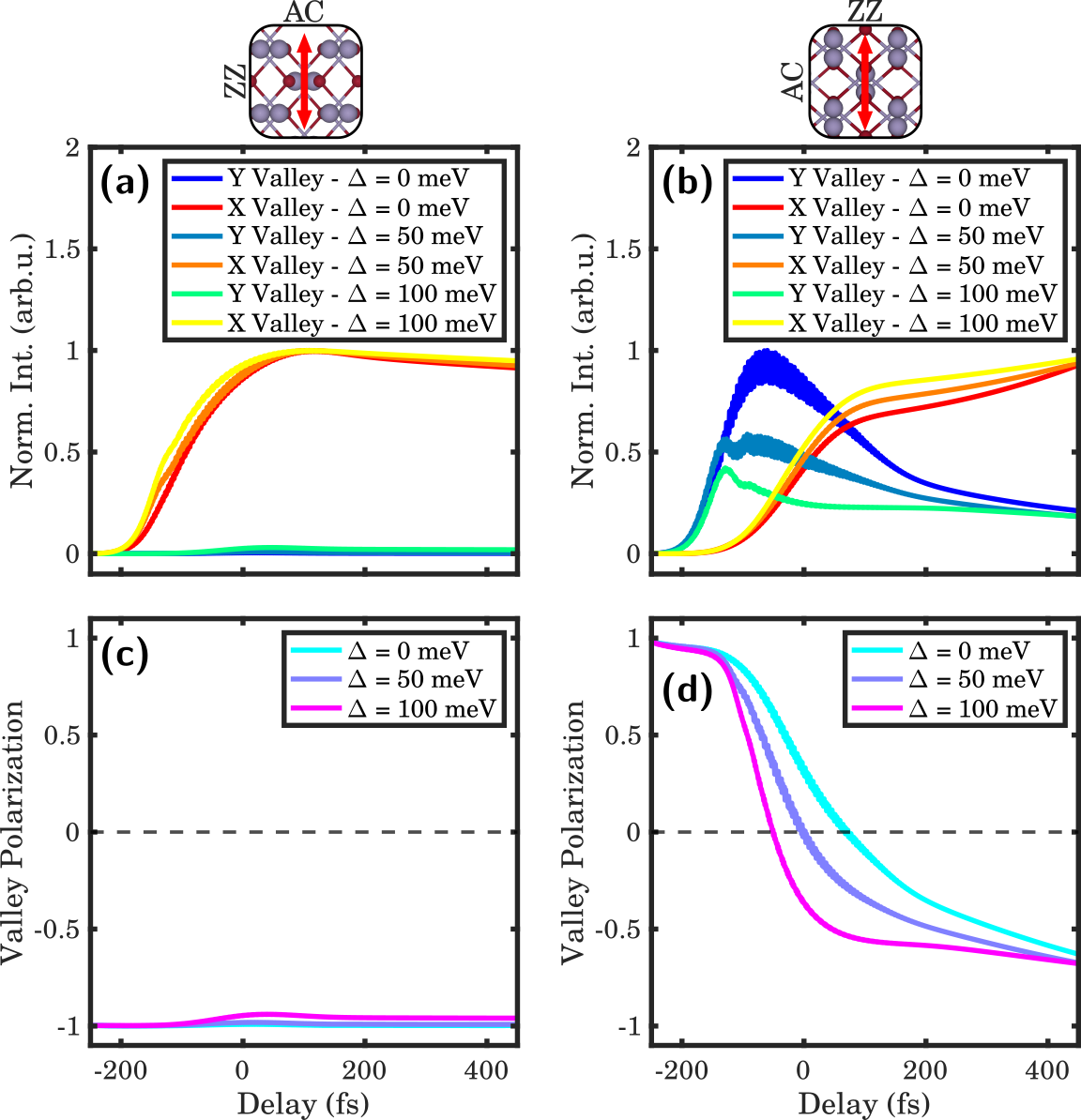}
\caption{\textbf{Role of detuning in time-dependent Boltzmann equation simulations of polarization-dependent valley-resolved ultrafast electron dynamics in SnSe.} \textbf{(a)-(b)} Valley-resolved electron dynamics following photoexcitation using IR polarization along ZZ and AC direction, respectively, as a function of pump photon energy detuning from the band edge ($\Delta$) and \textbf{(c)-(d)} associated valley polarization dynamics extracted by taking the normalized difference between population within CB X and Y valleys shown in \textbf{(a)} and \textbf{(b)}, respectively .}
\label{fig:rt-detuning}
\end{center}
\end{figure}

In Fig.~\ref{fig:rt-detuning}(a)-(b), we show the valley-resolved electron dynamics following photoexcitation with IR polarization along the ZZ and AC directions, respectively, as a function of pump photon energy detuning from the band edge ($\Delta$), along with the corresponding valley polarization dynamics in (c)-(d). Upon photoexcitation along the ZZ direction, where valleys located at the global minimum of the conduction band are predominantly populated, both the valley-resolved electron dynamics and the associated valley polarization exhibit only a weak dependence on pump photon energy detuning.

In contrast, for photoexcitation along the AC direction, the valley-resolved electron dynamics and associated valley polarization are significantly more sensitive to pump photon energy detuning. While the initial (early rising edge of the pulse) and final (after $\sim$400~fs) valley polarization values converge for all investigated detuning values, the intermediate dynamics clearly differ. Specifically, the decay and reversal of valley polarization occur more rapidly with increasing pump photon energy. This indicates that the generation of electrons with higher excess energy in the Y valleys promotes faster scattering of electrons out of these valleys.

Nonetheless, the overall behavior of the polarization-dependent nonequilibrium valley dynamics remains consistent across detuning: under ZZ pumping, the population remains nearly time-independent and almost fully X-valley polarized, whereas under AC pumping, the initially Y-valley-polarized population decays and even reverses on ultrafast timescales.

\section{Real-time anisotropic electron-hole dynamics in \NoCaseChange{SnSe}}\label{sec:rt-eh-dyn}

\begin{figure}[H]
\begin{center}
\includegraphics[width=0.65\textwidth]{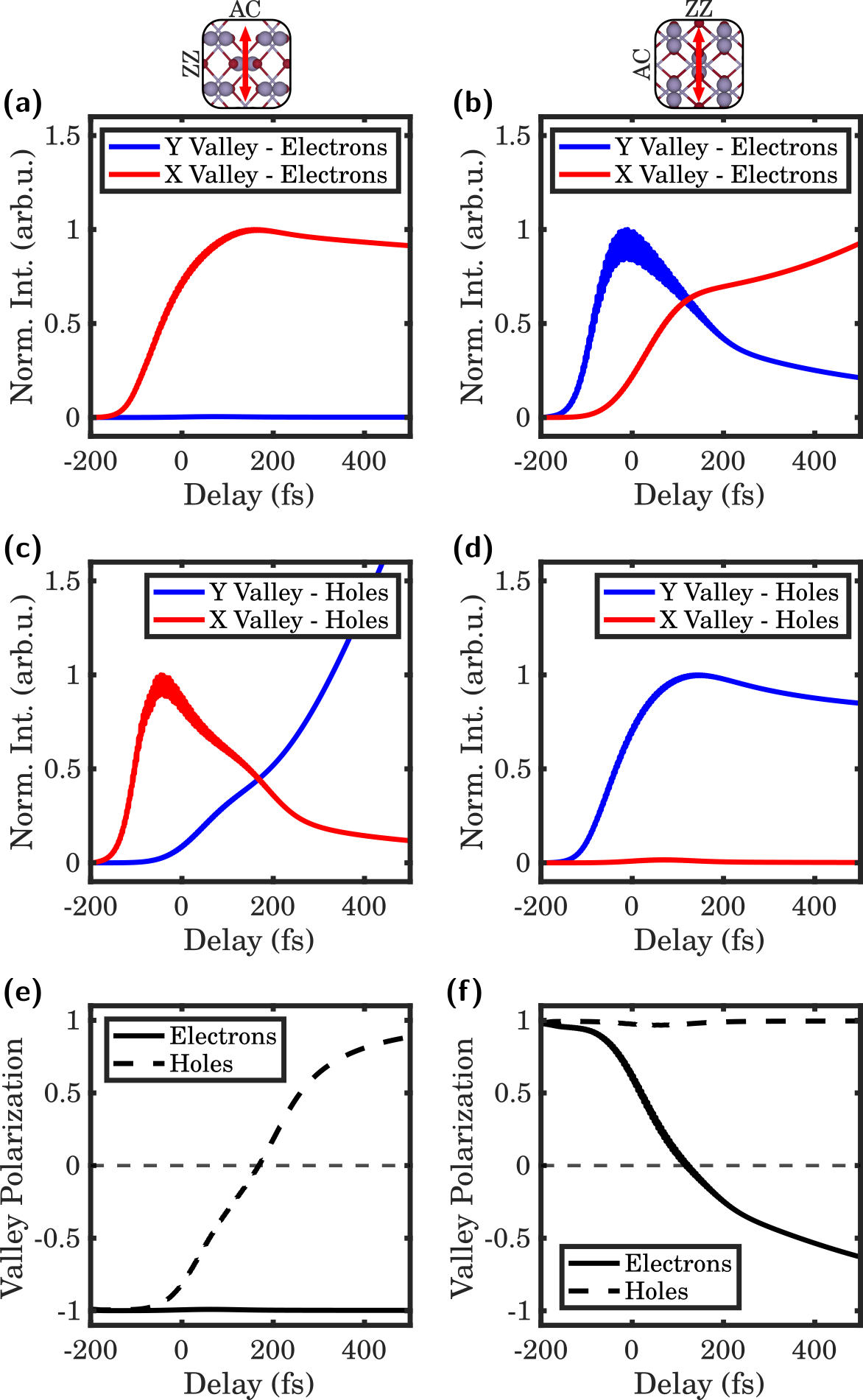}
\caption{\textbf{Time-dependent Boltzmann equation simulations of polarization-dependent valley-resolved ultrafast electron and hole dynamics in SnSe.} \textbf{(a)-(b)} Valley-resolved electron dynamics following photoexcitation using IR polarization along ZZ and AC direction, respectively, which are identical to the results presented in Figs. 2(e) and (f) of the main text. \textbf{(c)-(d)} Valley-resolved hole dynamics following photoexcitation using IR polarization along ZZ and AC direction, respectively. \textbf{(e)-(f)} Electrons and holes valley polarization dynamics.}
\label{fig:holes}
\end{center}
\end{figure}

In SnSe, the valence band maxima (VBM) lie at the Y valleys (along the AC direction), while the conduction band minima (CBM) are located at the X valleys (along the ZZ direction). We have demonstrated that this electronic structure dictates the strongly anisotropic ultrafast electron dynamics following polarization-controlled, valley-selective optical excitations. A natural follow-up question concerns the corresponding polarization- and valley-resolved dynamics of the photoholes. At early time delays, electrons and holes are expected to share the same valley polarization, since vertical interband optical transitions generate electron–hole pairs within the same valley. We have shown that, irrespective of the optical selection rules and the resulting initial valley polarization, the electronic population eventually relaxes to the global CBM (see Figs.~\ref{fig:holes}(a) and (b)), as expected for semiconductors. Conversely, at longer time delays, holes are anticipated to accumulate at the global VBM. Accordingly, when the optical transition occurs at the Y valleys (AC direction), where the VBM is located, a nearly unity and time-dependent hole valley polarization is observed (Fig.~\ref{fig:holes}(d) and (f)). In contrast, when the optical transition takes place at the X valleys (ZZ direction), we observe an initially strong negative valley polarization that subsequently decays and even reverses over time (Fig.~\ref{fig:holes}(c) and (e)). Because the VBM and CBM reside in opposite valleys, the ultrafast valley dynamics of electrons and holes are correspondingly opposite: the trend observed for electrons under AC photoexcitation mirrors that of holes under ZZ photoexcitation, and vice versa. 

These polarization- and valley-resolved electron and hole dynamics underscore the importance of a momentum-resolved probe for obtaining a complete understanding of nonequilibrium valley physics. Previous optical spectroscopic studies demonstrating selectivity in the absorption and emission of linearly polarized light, through optical absorption and photoluminescence measurements, are primarily sensitive to vertical interband transitions, i.e., processes in which electrons and holes reside in the same valleys. In our work, we show that, owing to the indirect band gap nature of SnSe, this condition is extremely short-lived: electrons and holes rapidly accumulate at the X (CBM) and Y (VBM) valleys, respectively, on a sub-picosecond timescale following valley-selective optical excitation. While optical spectroscopy serves as a powerful probe of the initial valley polarization, our approach provides a complete picture of nonequilibrium valley dynamics across the full energy–momentum landscape.

\section{Selection Rules for phonon-assisted intervalley scatterings.}
To understand the mode-resolved nonequilibrium phonon distribution from the TDBE calculations, we perform symmetry analysis of the e-ph coupling matrix elements involved in the intervalley scattering, which by definition is $g_{mn\nu}(\bk, \bq)=\frac{1}{\sqrt{2\omega_{\bq\nu}}}\bra{m\bk+\bq}\partial_{\bq\nu}V^{\rm scf}\ket{n\bk}$.

Only phonon modes with nonvanishing e-ph matrix elements can be emitted in the intervalley scattering process. This means that the product of the three components in the expression: the initial and final electronic wave functions and the phonon-induced potential must be invariant under all symmetry operations that are common to the little group of the three wavectors: $\bk$, $\bq$, and $\bk+\bq$.

The space group of bulk SnSe is $Pnma$ ($D_{2h}^{16}$ in Sch\"onflies notation), where the rotation parts of the symmetry operation are isomorphic to the point group $D_{2h}$, containing 8 symmetry operations and are listed in Table~\ref{D_2h}. At a finite wavevector $\bk$, part of these symmetries are broken and the remaining symmetries form a subgroup of $D_{2h}$, namely the little group of this wavevector. For the wave-vector $\bk = (k_x,0,0)$ along the $\rm \Gamma-X$, the wavevector is invariant under the symmetry operators no. 1, 3, 6 and 8, and the wavevector $\bk = (0,k_y,0)$ along the $\rm \Gamma-Y$ is invariant under the symmetries no. 1, 2, 7 and 8, both are isomorphic to $C_{2v}$ group. The wave-vector of the phonon $\bq=(q_x,q_y,0)$,  which is along $\rm \Gamma-M$ point, is only invariant under symmetries no. 1 and 8, which is equivalent to $C_s$ group. As a result, the subgroup of interest is $C_s$ , making of identity (no. 1) and gliding symmetry (no. 8) on a plane parallel to the $x-y$ plane.

By analysing the parity of the wave functions of the lowest conduction band at X- and Y-valleys, we found that they are both even under glide symmetry. As a result, the eigenvector of the phonon emitted must be even under gliding symmetry, which is satisfied by the second-highest-frequency phonon branch along $\rm \Gamma-M$ direction. 

\begin{figure}[H]
\begin{center}
\includegraphics[width=0.75\textwidth]{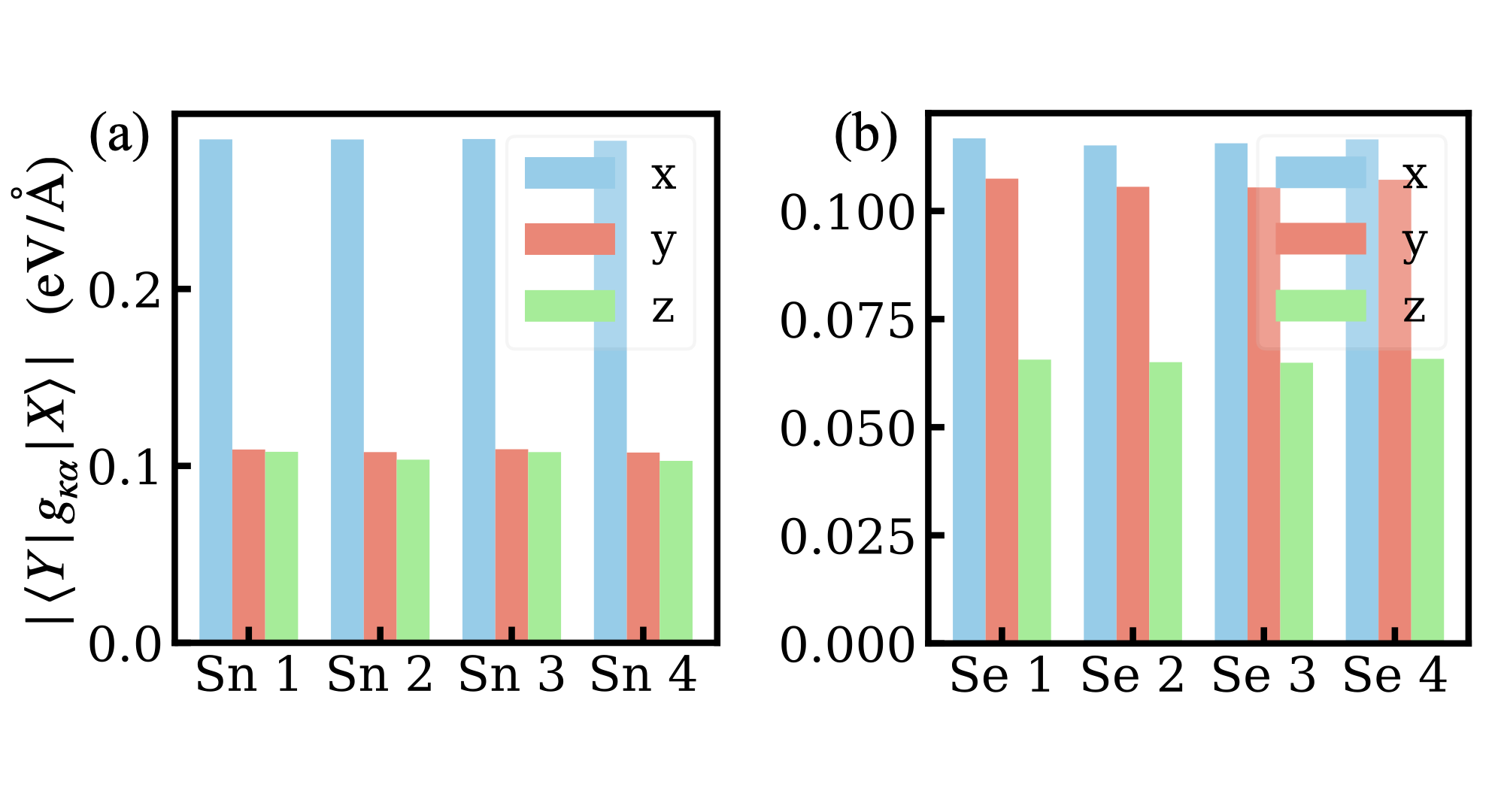}
\caption{Electron-phonon coupling between the two valleys of the conduction bands at X and Y projected on atomic displacements for the 4 Sn (a) and Se (b) atoms in the unit cell. }
\label{fig:g_atm}
\end{center}
\end{figure}

Symmetry analysis only explains qualitatively which phonons of the 24 modes are allowed, but does not provide a quantitative comparison of different modes that are even under gliding symmetry. In order to understand this, we perform orbital analysis of the wave functions and compare them to the displacement patterns of phonons.
\begin{table}
		\centering
		\caption{Symmetry operations of the D$_{2h}$ point group.}
		\label{D_2h}
		\begin{tabular}{|l|l|l|}
			\toprule
			\textbf{Symmetry} & \textbf{Symbol} & \textbf{Operation on wave vectors} \\ 
			\midrule
			1                  & $E$              & $(k_x, k_y, k_z) \rightarrow (k_x, k_y, k_z)$               \\
			
			2     & $\bar{C}_{2}(y)$       & $(k_x, k_y, k_z) \rightarrow (-k_x, k_y, -k_z)$              \\
			3     & $\bar{C}_{2}(x)$       & $(k_x, k_y, k_z) \rightarrow (k_x, -k_y, -k_z)$              \\
			4     & $\bar{C}_{2}(z)$       & $(k_x, k_y, k_z) \rightarrow (-k_x, -k_y, k_z)$              \\			
			5           & $i$              & $(k_x, k_y, k_z) \rightarrow (-k_x, -k_y, -k_z)$  \\
			6               & $\bar{M}_{xz}$    & $(k_x, k_y, k_z) \rightarrow (k_x, -k_y, k_z)$ \\
			7               & $\bar{M}_{yz}$    & $(k_x, k_y, k_z) \rightarrow (-k_x, k_y, k_z)$  \\
			8               & $\bar{M}_{xy}$    & $(k_x, k_y, k_z) \rightarrow (k_x, k_y, -k_z)$  \\
			\bottomrule
		\end{tabular}
		
		\vspace{0.5em}
	\end{table}
To identify the factors behind the mode selectivity, it is helpful to rewrite the expression of e-ph coupling matrix elements~\cite{giustino07}

\begin{align}
    g_{mn\nu}(\bk,\bq)&=\sum_{\kappa\alpha}\frac{e^{\alpha}_{\bq\nu,\kappa}}{\sqrt{2\omega_{\bq\nu}M_\kappa}} \cdot \sum_{p}{\rm exp}(i\bq\cdot\bR_p)\bra{m\bk+\bq}\frac{\partial V^{\rm scf}(\br)}{\partial\Delta\tau_{p\kappa\alpha}}\ket{n\bk}\nonumber \\  &=\sum_{\kappa\alpha}\frac{e^{\alpha}_{\bq\nu,\kappa}}{\sqrt{2\omega_{\bq\nu}M_\kappa}} g_{mn,\kappa\alpha}(\bk,\bq) \quad,
\end{align}

where ${\bf e}_{\bq\nu}$ represents the eigenvector of phonon mode $(\bq\nu)$, with the dimension of $3\times N_{at}$, and $e_{\bq\nu,\kappa}^{\alpha}$ is the component on atom $\kappa$ along the $\alpha$ direction. By extracting the projection of e-ph coupling matrix elements between the two valleys at CBM, i.e. $\ket{m\bk+\bq} = \ket{X}$ and $\ket{n\bk} = \ket{Y}$, to the atomic displacements $\kappa\alpha$, one can identify for the scattering between a certain pair of initial and final state, which atomic displacement contributes to the scattering. In Fig.\ref{fig:g_atm} we present the absolute value of each component of $g_{mn,\kappa\alpha}(\bk,\bq)$, from which we find that the intervalley scattering is more sensitive to the variation of self-consistent potential by in-plane displacement of the atoms. 

\clearpage

%\bibliography{References.bib} 

\begin{thebibliography}{10}
\expandafter\ifx\csname url\endcsname\relax
  \def\url#1{\texttt{#1}}\fi
\expandafter\ifx\csname urlprefix\endcsname\relax\def\urlprefix{URL }\fi
\providecommand{\bibinfo}[2]{#2}
\providecommand{\eprint}[2][]{\url{#2}}

\bibitem{PhysRevLett.108.126804}
\bibinfo{author}{Culcer, D.}, \bibinfo{author}{Saraiva, A.~L.}, \bibinfo{author}{Koiller, B.}, \bibinfo{author}{Hu, X.} \& \bibinfo{author}{Das~Sarma, S.}
\newblock Valley-Based Noise-Resistant Quantum Computation Using Si Quantum Dots.
\newblock \emph{\bibinfo{journal}{Phys. Rev. Lett.}} \textbf{\bibinfo{volume}{108}}, \bibinfo{pages}{126804} (\bibinfo{year}{2012}).
\newblock \urlprefix\url{https://link.aps.org/doi/10.1103/PhysRevLett.108.126804}.

\bibitem{Rohling_2012}
\bibinfo{author}{Rohling, N.} \& \bibinfo{author}{Burkard, G.}
\newblock Universal quantum computing with spin and valley states.
\newblock \emph{\bibinfo{journal}{New J. Phys.}} \textbf{\bibinfo{volume}{14}}, \bibinfo{pages}{083008} (\bibinfo{year}{2012}).
\newblock \urlprefix\url{https://doi.org/10.1088/1367-2630/14/8/083008}.

\bibitem{Schaibley16}
\bibinfo{author}{Schaibley, J.~R.} \emph{et~al.}
\newblock Valleytronics in 2D materials.
\newblock \emph{\bibinfo{journal}{Nat. Rev. Mater.}} \textbf{\bibinfo{volume}{1}}, \bibinfo{pages}{16055} (\bibinfo{year}{2016}).
\newblock \urlprefix\url{https://doi.org/10.1038/natrevmats.2016.55}.

\bibitem{Xiao12}
\bibinfo{author}{Xiao, D.}, \bibinfo{author}{Liu, G.-B.}, \bibinfo{author}{Feng, W.}, \bibinfo{author}{Xu, X.} \& \bibinfo{author}{Yao, W.}
\newblock Coupled Spin and Valley Physics in Monolayers of ${\mathrm{MoS}}_{2}$ and Other Group-VI Dichalcogenides.
\newblock \emph{\bibinfo{journal}{Phys. Rev. Lett.}} \textbf{\bibinfo{volume}{108}}, \bibinfo{pages}{196802} (\bibinfo{year}{2012}).
\newblock \urlprefix\url{https://link.aps.org/doi/10.1103/PhysRevLett.108.196802}.

\bibitem{XiaoDiWang2007}
\bibinfo{author}{Xiao, D.}, \bibinfo{author}{Yao, W.} \& \bibinfo{author}{Niu, Q.}
\newblock Valley-Contrasting Physics in Graphene: Magnetic Moment and Topological Transport.
\newblock \emph{\bibinfo{journal}{Phys. Rev. Lett.}} \textbf{\bibinfo{volume}{99}}, \bibinfo{pages}{236809} (\bibinfo{year}{2007}).
\newblock \urlprefix\url{https://link.aps.org/doi/10.1103/PhysRevLett.99.236809}.

\bibitem{Yao08}
\bibinfo{author}{Yao, W.}, \bibinfo{author}{Xiao, D.} \& \bibinfo{author}{Niu, Q.}
\newblock Valley-dependent optoelectronics from inversion symmetry breaking.
\newblock \emph{\bibinfo{journal}{Phys. Rev. B}} \textbf{\bibinfo{volume}{77}}, \bibinfo{pages}{235406} (\bibinfo{year}{2008}).
\newblock \urlprefix\url{https://link.aps.org/doi/10.1103/PhysRevB.77.235406}.

\bibitem{Mak12}
\bibinfo{author}{Mak, K.~F.}, \bibinfo{author}{He, K.}, \bibinfo{author}{Shan, J.} \& \bibinfo{author}{Heinz, T.~F.}
\newblock Control of valley polarization in monolayer MoS$_2$ by optical helicity.
\newblock \emph{\bibinfo{journal}{Nat. Nanotechnol.}} \textbf{\bibinfo{volume}{7}}, \bibinfo{pages}{494--498} (\bibinfo{year}{2012}).
\newblock \urlprefix\url{https://doi.org/10.1038/nnano.2012.96}.

\bibitem{Zeng12}
\bibinfo{author}{Zeng, H.}, \bibinfo{author}{Dai, J.}, \bibinfo{author}{Yao, W.}, \bibinfo{author}{Xiao, D.} \& \bibinfo{author}{Cui, X.}
\newblock Valley polarization in MoS$_2$ monolayers by optical pumping.
\newblock \emph{\bibinfo{journal}{Nat. Nanotechnol.}} \textbf{\bibinfo{volume}{7}}, \bibinfo{pages}{490--493} (\bibinfo{year}{2012}).
\newblock \urlprefix\url{https://doi.org/10.1038/nnano.2012.95}.

\bibitem{Cao12}
\bibinfo{author}{Cao, T.} \emph{et~al.}
\newblock Valley-selective circular dichroism of monolayer molybdenum disulphide.
\newblock \emph{\bibinfo{journal}{Nat. Commun.}} \textbf{\bibinfo{volume}{3}}, \bibinfo{pages}{887} (\bibinfo{year}{2012}).
\newblock \urlprefix\url{https://doi.org/10.1038/ncomms1882}.

\bibitem{sciadv.adk3897}
\bibinfo{author}{Beaulieu, S.} \emph{et~al.}
\newblock Berry curvature signatures in chiroptical excitonic transitions.
\newblock \emph{\bibinfo{journal}{Sci. Adv.}} \textbf{\bibinfo{volume}{10}}, \bibinfo{pages}{eadk3897} (\bibinfo{year}{2024}).
\newblock \urlprefix\url{https://www.science.org/doi/abs/10.1126/sciadv.adk3897}.

\bibitem{Fragkos2025_NC}
\bibinfo{author}{Fragkos, S.} \emph{et~al.}
\newblock Floquet-Bloch valleytronics.
\newblock \emph{\bibinfo{journal}{Nat. Commun.}} \textbf{\bibinfo{volume}{16}}, \bibinfo{pages}{5799} (\bibinfo{year}{2025}).
\newblock \urlprefix\url{https://doi.org/10.1038/s41467-025-61076-7}.

\bibitem{Zhu25}
\bibinfo{author}{Zhu, X.} \emph{et~al.}
\newblock A holistic view of the dynamics of long-lived valley polarized dark excitonic states in monolayer {WS}$_2$.
\newblock \emph{\bibinfo{journal}{Nat. Commun.}} \textbf{\bibinfo{volume}{16}}, \bibinfo{pages}{6385} (\bibinfo{year}{2025}).
\newblock \urlprefix\url{https://doi.org/10.1038/s41467-025-61677-2}.

\bibitem{Yu2014_DiracCones}
\bibinfo{author}{Yu, H.}, \bibinfo{author}{Liu, G.-B.}, \bibinfo{author}{Gong, P.}, \bibinfo{author}{Xu, X.} \& \bibinfo{author}{Yao, W.}
\newblock Dirac cones and Dirac saddle points of bright excitons in monolayer transition metal dichalcogenides.
\newblock \emph{\bibinfo{journal}{Nat. Commun.}} \textbf{\bibinfo{volume}{5}}, \bibinfo{pages}{3876} (\bibinfo{year}{2014}).
\newblock \urlprefix\url{https://doi.org/10.1038/ncomms4876}.

\bibitem{YuWu2014_ValleyDepolarization}
\bibinfo{author}{Yu, T.} \& \bibinfo{author}{Wu, M.~W.}
\newblock Valley depolarization due to intervalley and intravalley electron-hole exchange interactions in monolayer ${\text{MoS}}_{2}$.
\newblock \emph{\bibinfo{journal}{Phys. Rev. B}} \textbf{\bibinfo{volume}{89}}, \bibinfo{pages}{205303} (\bibinfo{year}{2014}).
\newblock \urlprefix\url{https://link.aps.org/doi/10.1103/PhysRevB.89.205303}.

\bibitem{Glazov2015_SpinValley}
\bibinfo{author}{Glazov, M.~M.} \emph{et~al.}
\newblock Spin and valley dynamics of excitons in transition metal dichalcogenide monolayers.
\newblock \emph{\bibinfo{journal}{Phys. Stat. Sol. b}} \textbf{\bibinfo{volume}{252}}, \bibinfo{pages}{2349} (\bibinfo{year}{2015}).
\newblock \urlprefix\url{https://onlinelibrary.wiley.com/doi/abs/10.1002/pssb.201552211}.

\bibitem{Yu2015_ValleyExcitons}
\bibinfo{author}{Yu, H.}, \bibinfo{author}{Cui, X.}, \bibinfo{author}{Xu, X.} \& \bibinfo{author}{Yao, W.}
\newblock Valley excitons in two-dimensional semiconductors.
\newblock \emph{\bibinfo{journal}{Nat. Sci. Rev.}} \textbf{\bibinfo{volume}{2}}, \bibinfo{pages}{57} (\bibinfo{year}{2015}).
\newblock \urlprefix\url{https://doi.org/10.1093/nsr/nwu078}.

\bibitem{Selig16}
\bibinfo{author}{Selig, M.} \emph{et~al.}
\newblock Excitonic linewidth and coherence lifetime in monolayer transition metal dichalcogenides.
\newblock \emph{\bibinfo{journal}{Nat. Commun.}} \textbf{\bibinfo{volume}{7}}, \bibinfo{pages}{13279} (\bibinfo{year}{2016}).
\newblock \urlprefix\url{https://doi.org/10.1038/ncomms13279}.

\bibitem{Bertoni16}
\bibinfo{author}{Bertoni, R.} \emph{et~al.}
\newblock Generation and Evolution of Spin-, Valley-, and Layer-Polarized Excited Carriers in Inversion-Symmetric ${\mathrm{WSe}}_{2}$.
\newblock \emph{\bibinfo{journal}{Phys. Rev. Lett.}} \textbf{\bibinfo{volume}{117}}, \bibinfo{pages}{277201} (\bibinfo{year}{2016}).
\newblock \urlprefix\url{https://link.aps.org/doi/10.1103/PhysRevLett.117.277201}.

\bibitem{Xu2021}
\bibinfo{author}{Xu, S.}, \bibinfo{author}{Si, C.}, \bibinfo{author}{Li, Y.}, \bibinfo{author}{Gu, B.-L.} \& \bibinfo{author}{Duan, W.}
\newblock Valley Depolarization Dynamics in Monolayer Transition-Metal Dichalcogenides: Role of the Satellite Valley.
\newblock \emph{\bibinfo{journal}{Nano Lett.}} \textbf{\bibinfo{volume}{21}}, \bibinfo{pages}{1785--1791} (\bibinfo{year}{2021}).
\newblock \urlprefix\url{https://doi.org/10.1021/acs.nanolett.0c04670}.

\bibitem{Molina17}
\bibinfo{author}{Molina-Sánchez, A.}, \bibinfo{author}{Sangalli, D.}, \bibinfo{author}{Wirtz, L.} \& \bibinfo{author}{Marini, A.}
\newblock Ab Initio Calculations of Ultrashort Carrier Dynamics in Two-Dimensional Materials: Valley Depolarization in Single-Layer WSe$_2$.
\newblock \emph{\bibinfo{journal}{Nano Lett.}} \textbf{\bibinfo{volume}{17}}, \bibinfo{pages}{4549--4555} (\bibinfo{year}{2017}).
\newblock \urlprefix\url{https://doi.org/10.1021/acs.nanolett.7b00175}.

\bibitem{MaiBarrette14}
\bibinfo{author}{Mai, C.} \emph{et~al.}
\newblock Many-Body Effects in Valleytronics: Direct Measurement of Valley Lifetimes in Single-Layer MoS$_2$.
\newblock \emph{\bibinfo{journal}{Nano Lett.}} \textbf{\bibinfo{volume}{14}}, \bibinfo{pages}{202--206} (\bibinfo{year}{2014}).
\newblock \urlprefix\url{https://doi.org/10.1021/nl403742j}.

\bibitem{ws2_valley_lifetime}
\bibinfo{author}{Beyer, H.} \emph{et~al.}
\newblock 80\% Valley Polarization of Free Carriers in Singly Oriented Single-Layer ${\mathrm{WS}}_{2}$ on Au(111).
\newblock \emph{\bibinfo{journal}{Phys. Rev. Lett.}} \textbf{\bibinfo{volume}{123}}, \bibinfo{pages}{236802} (\bibinfo{year}{2019}).
\newblock \urlprefix\url{https://link.aps.org/doi/10.1103/PhysRevLett.123.236802}.

\bibitem{PanCaruso2023}
\bibinfo{author}{Pan, Y.} \& \bibinfo{author}{Caruso, F.}
\newblock Vibrational Dichroism of Chiral Valley Phonons.
\newblock \emph{\bibinfo{journal}{Nano Lett.}} \textbf{\bibinfo{volume}{23}}, \bibinfo{pages}{7463--7469} (\bibinfo{year}{2023}).
\newblock \urlprefix\url{https://doi.org/10.1021/acs.nanolett.3c01904}.

\bibitem{ChenLi2025}
\bibinfo{author}{Chen, L.}, \bibinfo{author}{Li, Z.}, \bibinfo{author}{Li, Q.}, \bibinfo{author}{Zheng, Q.} \& \bibinfo{author}{Zhao, J.}
\newblock Spin Valley Dynamics Entangled with Optical Fields, Phonons, and Spin-Orbit Coupling in Monolayer {MoSe}$_2$.
\newblock \emph{\bibinfo{journal}{Adv. Optical Mater.}} \textbf{\bibinfo{volume}{13}}, \bibinfo{pages}{2403069} (\bibinfo{year}{2025}).
\newblock \urlprefix\url{https://advanced.onlinelibrary.wiley.com/doi/abs/10.1002/adom.202403069}.

\bibitem{PhysRevResearch.2.033340}
\bibinfo{author}{Wang, S.}, \bibinfo{author}{Ukhtary, M.~S.} \& \bibinfo{author}{Saito, R.}
\newblock Strain effect on circularly polarized electroluminescence in transition metal dichalcogenides.
\newblock \emph{\bibinfo{journal}{Phys. Rev. Res.}} \textbf{\bibinfo{volume}{2}}, \bibinfo{pages}{033340} (\bibinfo{year}{2020}).
\newblock \urlprefix\url{https://link.aps.org/doi/10.1103/PhysRevResearch.2.033340}.

\bibitem{An2023}
\bibinfo{author}{An, Z.} \emph{et~al.}
\newblock Strain control of exciton and trion spin-valley dynamics in monolayer transition metal dichalcogenides.
\newblock \emph{\bibinfo{journal}{Phys. Rev. B}} \textbf{\bibinfo{volume}{108}}, \bibinfo{pages}{L041404} (\bibinfo{year}{2023}).
\newblock \urlprefix\url{https://link.aps.org/doi/10.1103/PhysRevB.108.L041404}.

\bibitem{PhysRevB.110.125420}
\bibinfo{author}{Lan, K.}, \bibinfo{author}{Xie, S.} \& \bibinfo{author}{Fu, J.}
\newblock Laser-field detuning assisted optimization of valley dynamics in monolayer ${\mathrm{WSe}}_{2}$.
\newblock \emph{\bibinfo{journal}{Phys. Rev. B}} \textbf{\bibinfo{volume}{110}}, \bibinfo{pages}{125420} (\bibinfo{year}{2024}).
\newblock \urlprefix\url{https://link.aps.org/doi/10.1103/PhysRevB.110.125420}.

\bibitem{doi:10.1126/science.aac7820}
\bibinfo{author}{Rivera, P.} \emph{et~al.}
\newblock Valley-polarized exciton dynamics in a 2D semiconductor heterostructure.
\newblock \emph{\bibinfo{journal}{Science}} \textbf{\bibinfo{volume}{351}}, \bibinfo{pages}{688--691} (\bibinfo{year}{2016}).
\newblock \urlprefix\url{https://www.science.org/doi/abs/10.1126/science.aac7820}.

\bibitem{Zhang2022}
\bibinfo{author}{Zhang, Q.} \emph{et~al.}
\newblock Prolonging valley polarization lifetime through gate-controlled exciton-to-trion conversion in monolayer molybdenum ditelluride.
\newblock \emph{\bibinfo{journal}{Nat. Commun.}} \textbf{\bibinfo{volume}{13}}, \bibinfo{pages}{4101} (\bibinfo{year}{2022}).
\newblock \urlprefix\url{https://doi.org/10.1038/s41467-022-31672-y}.

\bibitem{PhysRevLett.114.037401}
\bibinfo{author}{MacNeill, D.} \emph{et~al.}
\newblock Breaking of Valley Degeneracy by Magnetic Field in Monolayer ${\mathrm{MoSe}}_{2}$.
\newblock \emph{\bibinfo{journal}{Phys. Rev. Lett.}} \textbf{\bibinfo{volume}{114}}, \bibinfo{pages}{037401} (\bibinfo{year}{2015}).
\newblock \urlprefix\url{https://link.aps.org/doi/10.1103/PhysRevLett.114.037401}.

\bibitem{PhysRevB.97.041405}
\bibinfo{author}{Xu, L.} \emph{et~al.}
\newblock Large valley splitting in monolayer ${\mathbf{WS}}_{\mathbf{2}}$ by proximity coupling to an insulating antiferromagnetic substrate.
\newblock \emph{\bibinfo{journal}{Phys. Rev. B}} \textbf{\bibinfo{volume}{97}}, \bibinfo{pages}{041405} (\bibinfo{year}{2018}).
\newblock \urlprefix\url{https://link.aps.org/doi/10.1103/PhysRevB.97.041405}.

\bibitem{Gindl25}
\bibinfo{author}{Gindl, A.}, \bibinfo{author}{{\v{C}}mel, M.}, \bibinfo{author}{Troj{\'a}nek, F.}, \bibinfo{author}{Mal{\'y}, P.} \& \bibinfo{author}{Koz{\'a}k, M.}
\newblock Ultrafast room-temperature valley manipulation in silicon and diamond.
\newblock \emph{\bibinfo{journal}{Nat. Phys.}} \textbf{\bibinfo{volume}{21}}, \bibinfo{pages}{947--952} (\bibinfo{year}{2025}).
\newblock \urlprefix\url{https://doi.org/10.1038/s41567-025-02862-4}.

\bibitem{Barraza21}
\bibinfo{author}{Barraza-Lopez, S.}, \bibinfo{author}{Fregoso, B.~M.}, \bibinfo{author}{Villanova, J.~W.}, \bibinfo{author}{Parkin, S. S.~P.} \& \bibinfo{author}{Chang, K.}
\newblock Colloquium: Physical properties of group-IV monochalcogenide monolayers.
\newblock \emph{\bibinfo{journal}{Rev. Mod. Phys.}} \textbf{\bibinfo{volume}{93}}, \bibinfo{pages}{011001} (\bibinfo{year}{2021}).
\newblock \urlprefix\url{https://link.aps.org/doi/10.1103/RevModPhys.93.011001}.

\bibitem{Rodin16}
\bibinfo{author}{Rodin, A.~S.}, \bibinfo{author}{Gomes, L.~C.}, \bibinfo{author}{Carvalho, A.} \& \bibinfo{author}{Castro~Neto, A.~H.}
\newblock Valley physics in tin (II) sulfide.
\newblock \emph{\bibinfo{journal}{Phys. Rev. B}} \textbf{\bibinfo{volume}{93}}, \bibinfo{pages}{045431} (\bibinfo{year}{2016}).
\newblock \urlprefix\url{https://link.aps.org/doi/10.1103/PhysRevB.93.045431}.

\bibitem{Lin2018}
\bibinfo{author}{Lin, S.} \emph{et~al.}
\newblock Accessing valley degree of freedom in bulk Tin(II) sulfide at room temperature.
\newblock \emph{\bibinfo{journal}{Nat. Commun.}} \textbf{\bibinfo{volume}{9}}, \bibinfo{pages}{1455} (\bibinfo{year}{2018}).
\newblock \urlprefix\url{https://doi.org/10.1038/s41467-018-03897-3}.

\bibitem{Chen18}
\bibinfo{author}{Chen, C.} \emph{et~al.}
\newblock Valley-Selective Linear Dichroism in Layered Tin Sulfide.
\newblock \emph{\bibinfo{journal}{ACS Photonics}} \textbf{\bibinfo{volume}{5}}, \bibinfo{pages}{3814--3819} (\bibinfo{year}{2018}).
\newblock \urlprefix\url{https://doi.org/10.1021/acsphotonics.8b00850}.

\bibitem{Tien24}
\bibinfo{author}{Thanh~Tien, N.}, \bibinfo{author}{Thi Bich~Thao, P.}, \bibinfo{author}{Thi~Han, N.} \& \bibinfo{author}{Khuong~Dien, V.}
\newblock Symmetry-driven valleytronics in the single-layer tin chalcogenides SnS and SnSe.
\newblock \emph{\bibinfo{journal}{Phys. Rev. B}} \textbf{\bibinfo{volume}{109}}, \bibinfo{pages}{155416} (\bibinfo{year}{2024}).
\newblock \urlprefix\url{https://link.aps.org/doi/10.1103/PhysRevB.109.155416}.

\bibitem{hashmi25}
\bibinfo{author}{Hashmi, A.} \emph{et~al.}
\newblock Ultrafast optical control of multivalley states in two-dimensional SnS.
\newblock \emph{\bibinfo{journal}{Phys. Rev. Mater.}} \textbf{\bibinfo{volume}{9}}, \bibinfo{pages}{104003} (\bibinfo{year}{2025}).
\newblock \urlprefix\url{https://link.aps.org/doi/10.1103/jm56-qldh}.

\bibitem{Hanakata2016}
\bibinfo{author}{Hanakata, P.~Z.}, \bibinfo{author}{Carvalho, A.}, \bibinfo{author}{Campbell, D.~K.} \& \bibinfo{author}{Park, H.~S.}
\newblock Polarization and valley switching in monolayer group-IV monochalcogenides.
\newblock \emph{\bibinfo{journal}{Phys. Rev. B}} \textbf{\bibinfo{volume}{94}}, \bibinfo{pages}{035304} (\bibinfo{year}{2016}).
\newblock \urlprefix\url{https://link.aps.org/doi/10.1103/PhysRevB.94.035304}.

\bibitem{Tolloczko25}
\bibinfo{author}{Tołłoczko, A.~K.} \emph{et~al.}
\newblock Linear Dichroism of the Optical Properties of SnS and SnSe Van der Waals Crystals.
\newblock \emph{\bibinfo{journal}{Small}} \textbf{\bibinfo{volume}{21}}, \bibinfo{pages}{2410903} (\bibinfo{year}{2025}).
\newblock \urlprefix\url{https://onlinelibrary.wiley.com/doi/abs/10.1002/smll.202410903}.

\bibitem{Fragkos2025}
\bibinfo{author}{Fragkos, S.} \emph{et~al.}
\newblock Time- and polarization-resolved extreme ultraviolet momentum microscopy.
\newblock \emph{\bibinfo{journal}{Review of Scientific Instruments}} \textbf{\bibinfo{volume}{96}}, \bibinfo{pages}{115201} (\bibinfo{year}{2025}).
\newblock \urlprefix\url{https://doi.org/10.1063/5.0260193}.

\bibitem{Comby22}
\bibinfo{author}{Comby, A.} \emph{et~al.}
\newblock Ultrafast polarization-tunable monochromatic extreme ultraviolet source at high-repetition-rate.
\newblock \emph{\bibinfo{journal}{Journal of Optics}} \textbf{\bibinfo{volume}{24}}, \bibinfo{pages}{084003} (\bibinfo{year}{2022}).
\newblock \urlprefix\url{https://dx.doi.org/10.1088/2040-8986/ac7a49}.

\bibitem{Medjanik17}
\bibinfo{author}{Medjanik, K.} \emph{et~al.}
\newblock Direct 3D mapping of the Fermi surface and Fermi velocity.
\newblock \emph{\bibinfo{journal}{Nat. Mater.}} \textbf{\bibinfo{volume}{16}}, \bibinfo{pages}{615--621} (\bibinfo{year}{2017}).
\newblock \urlprefix\url{https://doi.org/10.1038/nmat4875}.

\bibitem{tkach24}
\bibinfo{author}{Tkach, O.} \& \bibinfo{author}{Schönhense, G.}
\newblock Multimode objective lens for momentum microscopy and XPEEM: Theory.
\newblock \emph{\bibinfo{journal}{Ultramicroscopy}} \textbf{\bibinfo{volume}{276}}, \bibinfo{pages}{114167} (\bibinfo{year}{2025}).
\newblock \urlprefix\url{https://www.sciencedirect.com/science/article/pii/S030439912500066X}.

\bibitem{tkach24-2}
\bibinfo{author}{Tkach, O.} \emph{et~al.}
\newblock Multi-Mode Front Lens for Momentum Microscopy: Part {II} Experiments (\bibinfo{year}{2024}).
\newblock \urlprefix\url{https://arxiv.org/abs/2401.10084}.
\newblock \eprint{2401.10084}.

\bibitem{caruso2022}
\bibinfo{author}{Caruso, F.} \& \bibinfo{author}{Novko, D.}
\newblock Ultrafast dynamics of electrons and phonons: from the two-temperature model to the time-dependent Boltzmann equation.
\newblock \emph{\bibinfo{journal}{Adv. Phys.: X}} \textbf{\bibinfo{volume}{7}}, \bibinfo{pages}{2095925} (\bibinfo{year}{2022}).
\newblock \urlprefix\url{https://doi.org/10.1080/23746149.2022.2095925}.

\bibitem{caruso21}
\bibinfo{author}{Caruso, F.}
\newblock Nonequilibrium Lattice Dynamics in Monolayer MoS$_2$.
\newblock \emph{\bibinfo{journal}{J. Phys. Chem. Lett.}} \textbf{\bibinfo{volume}{12}}, \bibinfo{pages}{1734} (\bibinfo{year}{2021}).
\newblock \urlprefix\url{https://doi.org/10.1021/acs.jpclett.0c03616}.

\bibitem{Suppl}
\bibinfo{note}{See Supplemental Material at [URL].}

\bibitem{Kunin23}
\bibinfo{author}{Kunin, A.} \emph{et~al.}
\newblock Momentum-Resolved Exciton Coupling and Valley Polarization Dynamics in Monolayer ${\mathrm{WS}}_{2}$.
\newblock \emph{\bibinfo{journal}{Phys. Rev. Lett.}} \textbf{\bibinfo{volume}{130}}, \bibinfo{pages}{046202} (\bibinfo{year}{2023}).
\newblock \urlprefix\url{https://link.aps.org/doi/10.1103/PhysRevLett.130.046202}.

\bibitem{CarusoTroppenz2018}
\bibinfo{author}{Caruso, F.}, \bibinfo{author}{Troppenz, M.}, \bibinfo{author}{Rigamonti, S.} \& \bibinfo{author}{Draxl, C.}
\newblock Thermally enhanced Fr\"ohlich coupling in SnSe.
\newblock \emph{\bibinfo{journal}{Phys. Rev. B}} \textbf{\bibinfo{volume}{99}}, \bibinfo{pages}{081104} (\bibinfo{year}{2019}).
\newblock \urlprefix\url{https://link.aps.org/doi/10.1103/PhysRevB.99.081104}.

\bibitem{Macneill15}
\bibinfo{author}{MacNeill, D.} \emph{et~al.}
\newblock Breaking of Valley Degeneracy by Magnetic Field in Monolayer ${\mathrm{MoSe}}_{2}$.
\newblock \emph{\bibinfo{journal}{Phys. Rev. Lett.}} \textbf{\bibinfo{volume}{114}}, \bibinfo{pages}{037401} (\bibinfo{year}{2015}).
\newblock \urlprefix\url{https://link.aps.org/doi/10.1103/PhysRevLett.114.037401}.

\bibitem{Qi15}
\bibinfo{author}{Qi, J.}, \bibinfo{author}{Li, X.}, \bibinfo{author}{Niu, Q.} \& \bibinfo{author}{Feng, J.}
\newblock Giant and tunable valley degeneracy splitting in ${\mathrm{MoTe}}_{2}$.
\newblock \emph{\bibinfo{journal}{Phys. Rev. B}} \textbf{\bibinfo{volume}{92}}, \bibinfo{pages}{121403} (\bibinfo{year}{2015}).
\newblock \urlprefix\url{https://link.aps.org/doi/10.1103/PhysRevB.92.121403}.

\bibitem{Nagler17}
\bibinfo{author}{Nagler, P.} \emph{et~al.}
\newblock Giant magnetic splitting inducing near-unity valley polarization in van der Waals heterostructures.
\newblock \emph{\bibinfo{journal}{Nat. Commun.}} \textbf{\bibinfo{volume}{8}}, \bibinfo{pages}{1551} (\bibinfo{year}{2017}).
\newblock \urlprefix\url{https://doi.org/10.1038/s41467-017-01748-1}.

\bibitem{Guo24}
\bibinfo{author}{Guo, S.-D.}, \bibinfo{author}{Liu, Y.}, \bibinfo{author}{Yu, J.} \& \bibinfo{author}{Liu, C.-C.}
\newblock Valley polarization in twisted altermagnetism.
\newblock \emph{\bibinfo{journal}{Phys. Rev. B}} \textbf{\bibinfo{volume}{110}}, \bibinfo{pages}{L220402} (\bibinfo{year}{2024}).
\newblock \urlprefix\url{https://link.aps.org/doi/10.1103/PhysRevB.110.L220402}.

\bibitem{Tong2016}
\bibinfo{author}{Tong, W.-Y.}, \bibinfo{author}{Gong, S.-J.}, \bibinfo{author}{Wan, X.} \& \bibinfo{author}{Duan, C.-G.}
\newblock Concepts of ferrovalley material and anomalous valley Hall effect.
\newblock \emph{\bibinfo{journal}{Nat. Commun.}} \textbf{\bibinfo{volume}{7}}, \bibinfo{pages}{13612} (\bibinfo{year}{2016}).
\newblock \urlprefix\url{https://doi.org/10.1038/ncomms13612}.

\bibitem{Xian20}
\bibinfo{author}{Xian, R.~P.} \emph{et~al.}
\newblock An open-source, end-to-end workflow for multidimensional photoemission spectroscopy.
\newblock \emph{\bibinfo{journal}{Scientific Data}} \textbf{\bibinfo{volume}{7}}, \bibinfo{pages}{442} (\bibinfo{year}{2020}).
\newblock \urlprefix\url{https://doi.org/10.1038/s41597-020-00769-8}.

\bibitem{Xian19_2}
\bibinfo{author}{Xian, R.~P.}, \bibinfo{author}{Rettig, L.} \& \bibinfo{author}{Ernstorfer, R.}
\newblock Symmetry-guided nonrigid registration: The case for distortion correction in multidimensional photoemission spectroscopy.
\newblock \emph{\bibinfo{journal}{Ultramicroscopy}} \textbf{\bibinfo{volume}{202}}, \bibinfo{pages}{133 -- 139} (\bibinfo{year}{2019}).
\newblock \urlprefix\url{http://www.sciencedirect.com/science/article/pii/S0304399118303474}.

\bibitem{KuhnRossi1992}
\bibinfo{author}{Kuhn, T.} \& \bibinfo{author}{Rossi, F.}
\newblock Monte Carlo simulation of ultrafast processes in photoexcited semiconductors: Coherent and incoherent dynamics.
\newblock \emph{\bibinfo{journal}{Phys. Rev. B}} \textbf{\bibinfo{volume}{46}}, \bibinfo{pages}{7496} (\bibinfo{year}{1992}).
\newblock \urlprefix\url{https://link.aps.org/doi/10.1103/PhysRevB.46.7496}.

\bibitem{Caruso22}
\bibinfo{author}{Caruso, F.}, \bibinfo{author}{Schebek, M.}, \bibinfo{author}{Pan, Y.}, \bibinfo{author}{Vona, C.} \& \bibinfo{author}{Draxl, C.}
\newblock Chirality of Valley Excitons in Monolayer Transition-Metal Dichalcogenides.
\newblock \emph{\bibinfo{journal}{J. Phys. Chem. Lett.}} \textbf{\bibinfo{volume}{13}}, \bibinfo{pages}{5894--5899} (\bibinfo{year}{2022}).
\newblock \urlprefix\url{https://doi.org/10.1021/acs.jpclett.2c01034}.

\bibitem{Noffsinger2010}
\bibinfo{author}{Noffsinger, J.} \emph{et~al.}
\newblock {EPW}: A program for calculating the electron--phonon coupling using maximally localized {Wannier} functions.
\newblock \emph{\bibinfo{journal}{Comput. Phys. Commun.}} \textbf{\bibinfo{volume}{181}}, \bibinfo{pages}{2140} (\bibinfo{year}{2010}).
\newblock \urlprefix\url{https://doi.org/10.1016/j.cpc.2010.08.027}.

\bibitem{giustino07}
\bibinfo{author}{Giustino, F.}, \bibinfo{author}{Cohen, M.~L.} \& \bibinfo{author}{Louie, S.~G.}
\newblock Electron-phonon interaction using Wannier functions.
\newblock \emph{\bibinfo{journal}{Phys. Rev. B}} \textbf{\bibinfo{volume}{76}}, \bibinfo{pages}{165108} (\bibinfo{year}{2007}).
\newblock \urlprefix\url{https://link.aps.org/doi/10.1103/PhysRevB.76.165108}.

\bibitem{Lee2023}
\bibinfo{author}{Lee, H.} \emph{et~al.}
\newblock Electron--phonon physics from first principles using the EPW code.
\newblock \emph{\bibinfo{journal}{npj Comput. Mater.}} \textbf{\bibinfo{volume}{9}}, \bibinfo{pages}{156} (\bibinfo{year}{2023}).
\newblock \urlprefix\url{https://www.nature.com/articles/s41524-023-01107-3}.

\bibitem{Giannozzi2017}
\bibinfo{author}{Giannozzi, P.} \emph{et~al.}
\newblock Advanced capabilities for materials modelling with \textsc{Quantum ESPRESSO}.
\newblock \emph{\bibinfo{journal}{J. Phys.: Condens. Matter}} \textbf{\bibinfo{volume}{29}}, \bibinfo{pages}{465901} (\bibinfo{year}{2017}).
\newblock \urlprefix\url{https://iopscience.iop.org/article/10.1088/1361-648X/aa8f79}.

\bibitem{Baroni2001}
\bibinfo{author}{Baroni, S.}, \bibinfo{author}{de~Gironcoli, S.}, \bibinfo{author}{Dal~Corso, A.} \& \bibinfo{author}{Giannozzi, P.}
\newblock Phonons and related crystal properties from density-functional perturbation theory.
\newblock \emph{\bibinfo{journal}{Rev. Mod. Phys.}} \textbf{\bibinfo{volume}{73}}, \bibinfo{pages}{515} (\bibinfo{year}{2001}).
\newblock \urlprefix\url{https://journals.aps.org/rmp/abstract/10.1103/RevModPhys.73.515}.

\bibitem{hgh_pseudopotential}
\bibinfo{author}{Hartwigsen, C.}, \bibinfo{author}{Goedecker, S.} \& \bibinfo{author}{Hutter, J.}
\newblock Relativistic separable dual-space Gaussian pseudopotentials from H to Rn.
\newblock \emph{\bibinfo{journal}{Phys. Rev. B}} \textbf{\bibinfo{volume}{58}}, \bibinfo{pages}{3641--3662} (\bibinfo{year}{1998}).
\newblock \urlprefix\url{https://link.aps.org/doi/10.1103/PhysRevB.58.3641}.

\bibitem{PBE_GGA}
\bibinfo{author}{Perdew, J.~P.}, \bibinfo{author}{Burke, K.} \& \bibinfo{author}{Ernzerhof, M.}
\newblock Generalized Gradient Approximation Made Simple.
\newblock \emph{\bibinfo{journal}{Phys. Rev. Lett.}} \textbf{\bibinfo{volume}{77}}, \bibinfo{pages}{3865--3868} (\bibinfo{year}{1996}).
\newblock \urlprefix\url{https://link.aps.org/doi/10.1103/PhysRevLett.77.3865}.

\bibitem{Marzari2012}
\bibinfo{author}{Marzari, N.}, \bibinfo{author}{Mostofi, A.~A.}, \bibinfo{author}{Yates, J.~R.}, \bibinfo{author}{Souza, I.} \& \bibinfo{author}{Vanderbilt, D.}
\newblock Maximally localized {Wannier} functions: Theory and applications.
\newblock \emph{\bibinfo{journal}{Rev. Mod. Phys.}} \textbf{\bibinfo{volume}{84}}, \bibinfo{pages}{1419} (\bibinfo{year}{2012}).
\newblock \urlprefix\url{https://doi.org/10.1103/RevModPhys.84.1419}.

\end{thebibliography}

\providecommand{\noopsort}[1]{}\providecommand{\singleletter}[1]{#1}%

\end{document}